\documentclass{article}
\usepackage[a4paper,top=2.54cm,bottom=2.0cm,left=2.0cm,right=2.54cm]{geometry}
\usepackage[english]{babel}
\usepackage[utf8]{inputenc}
\usepackage[T1]{fontenc}
\usepackage{courier}
\usepackage{helvet}
\usepackage{etoolbox}
\usepackage{graphicx}
\usepackage[font=small,format=plain,labelfont=bf,up,textfont=it,up]{caption}
\usepackage{fancyhdr}
\usepackage{booktabs}
\usepackage[usenames,svgnames,dvipsnames]{xcolor}
\usepackage[pdftex,plainpages=false,pdfpagelabels,colorlinks=true,citecolor=DarkGreen,linkcolor=DarkBlue,urlcolor=black,filecolor=green,bookmarksopen=true,pagebackref, hypertexnames=false]{hyperref}
\usepackage[all]{hypcap} 

\usepackage{titlesec}
\titleformat{\section}{\normalfont\fontfamily{lmr}\fontsize{13}{17}\bfseries}{\thesection}{1em}{}
\titleformat{\subsection}{\normalfont\fontfamily{lmr}\fontsize{11}{17}\itshape}{\thesubsection}{1em}{}
\titleformat{\author}{\normalfont\fontfamily{lmr}\fontsize{10}{15}\bfseries}{\thesection}{1em}{}

\usepackage{amssymb}
\usepackage{amsthm}
\usepackage{amsmath}
\usepackage{subfig}

\newtheorem{theorem}{Theorem}[section]

\newtheorem{lemma}[theorem]{Lemma}
\newtheorem{definition}[theorem]{Definition}
\newtheorem{remark}[theorem]{Remark}
\newtheorem{proposition}[theorem]{Proposition}

\usepackage{authblk}

\newsavebox\affbox
\author[1*]{Lorena C. Bulhosa}
\author[2*]{Juliane F. Oliveira}
\affil[1]{Institute for Pure and Applied Mathematics, 22460-320, Rio de Janeiro, Brazil
}
\affil[2]{Center for Data and Knowledge Integration for Health, Gonçalo Moniz Institute, Oswaldo Cruz Foundation, 40296-710 Salvador, Bahia, Brazil
}

\title{Vaccination in a two-strain model with cross-immunity and antibody-dependent enhancement\\
	}
\date{}    

\begin{document}

\setcounter{page}{1}
\renewcommand{\thepage}{\arabic{page}}	
\maketitle
	
\noindent\rule{15cm}{0.5pt}
	\begin{abstract}
		Dengue and Zika incidence data and the latest research have raised questions  about how dengue vaccine strategies might be impacted by the emergence of Zika virus. Existing antibodies to one virus might temporarily protect or promote infection by the other through antibody-dependent enhancement (ADE). With this condition, understanding the dynamics of propagation of these two viruses is of great importance when implementing vaccines. In this work, we analyze the effect of vaccination against one strain, in a two-strain model that accounts for cross-immunity and ADE. Using basic and invasion reproductive numbers, we examined the dynamics of the model and provide conditions to ensure the stability of the disease-free equilibrium. We provide conditions on cross-immunity, ADE and vaccination rate under which the vaccination could ensure the global stability of the disease-free equilibrium. The results indicate scenarios in which vaccination against one strain may improve or worsen the control of the other, as well as contribute to the eradication or persistence of one or both viruses in the population.  \\ \\
		\let\thefootnote\relax\footnotetext{
			\small $^{*}$\textbf{Corresponding authors.} \textit{
				\textit{E-mail address: lbulhosa@impa.br} (Lorena C. Bulhosa), julianlanzin@gmail.com (Juliane F. Oliveira)}		
		}
	\end{abstract}
\noindent\rule{15cm}{0.4pt}

\section{Introduction}
Dengue and Zika are two important arbovirus affecting humans. Global dengue incidence has increased dramatically, putting about half the world's population at risk. According to one estimate, around 100 to 400 million cases of dengue occur worldwide each year, resulting in around 20,000 deaths \cite{dengue-brasil,dengue-artigo,mondiale2018dengue}. Zika virus (ZIKV) became better known in 2016, when pregnant women pre-exposed to ZIKV infection caused disabilities and microcephaly in newborns. ZIKV has been detected in 89 countries, posing infected individuals at a higher risk for severe neurologic sequelae. Its infections cause symptoms similar to dengue disease, leading to massive misdiagnosis in co-spreading countries \cite{oliveira2020interdependence,opas-oms-zika}.

Planning prevention and control strategies to reduce the burden of dengue virus (DENV) and ZIKV is no easy task. Both viruses are mainly transmitted by the Aedes Aegypti mosquito, which is abundant in settings with environmental conditions favorable to its development and proliferation. In addition, DENV transmission is strongly influenced by the dynamic spread of its four serotypes, and it have been affected by the emergence of ZIKV \cite{dz-tropical}. The level of antibodies against one dengue serotype can cause different body reactions in the case of a secondary infection \cite{katzelnick}. Recovery from infection by a dengue serotype provides lifelong immunity to that sorotype. However, individuals who later become infected with a different serotype may experience \textit{antibody-dependent enhancement} (ADE), where antibodies from a previous infection do not protect (in a long term) against a new infection, but increase the individual susceptibility and the risk of severe outcomes \cite{dengue-ade}. DENV and ZIKV both come from the flavivirus family and are genetically similar. Therefore, the interactions between these viruses could be similar to those between two different dengue virus serotypes \cite{dz-science,dz-lancet}.

The issues surrounding the interactions between Zika and dengue fever have potential implications for case surveillance and vaccine development \cite{essink2023safety,dz-frontiers, who_dengue_vaccine,dengue-vacina}. As researchers continue to analyze the data and the biological basis of their findings, mathematical models provide a tool that can contribute to understand vaccination strategies and its  implications on the dynamics of multi-strain circulation \cite{billings,dengue-simulations,dengue-zika-vacina}. By assuming that the vaccine would be less effective against the serotype with the highest transmission intensity, the authors in \cite{dengue-simulations} found that vaccine may be effective against the weaker strain but contribute to an increase in incidence of secondary infections of the stronger one. However, in the long term, vaccination strategies could reduce the overall proportion of infections, but still with periodic yearly outbreaks of the strong strain.

In \cite{billings}, ADE effect was studied in a two-sorotype dengue model with vaccination against one and both strains. They concluded that if vaccination is against only one strain, eradication of the other will not be achieved if it was currently in an endemic state. If the population is vaccinated against both strains separately, there are conditions on vaccination rates to ensure the eradication of both diseases. But using dengue parameters, this strategy might not work if a person cannot receive both vaccines. Zika and dengue interaction was investigated in \cite{dengue-zika-vacina}. The authors constructed a vaccination model considering the ADE effect and the possibility of co-infection by both viruses. They analyzed the dynamics of the model through basic and invasion numbers. Their results show a positive vaccination effect in controlling dengue. However, their simulations indicate an increase in Zika incidence due to ADE.

Few works in the literature examine the full specificities of the interaction of dengue and Zika. In this work, we develop a more general approach by modelling the effect of dengue vaccination in a two-strain model, considering both temporary cross-immunity and the ADE effect between strains. After model formulation, we calculate the main equilibria and the basic and invasion reproductive numbers of both viruses. We study the dynamics of the model and provided conditions for the local and global stability of the main equilibria and the persistence of one or both diseases. Finally, the effect of the ADE factor and temporary cross-immunity are examined. Vaccination criteria are established and simulations are performed to illustrate the  possible outcomes of vaccination strategies.

\section{Model formulation}\label{sec:modelo}

The model describes the circulation of two strains, denoted $1$ and $2$, with a vaccination against strain $1$. Note that in the Dengue and Zika transmission scenario, the strains can represent the Dengue and Zika viruses. To simplify the model, the mosquito population is not taken into account.

The population is divided into groups: susceptible individuals to both strains, $S$; vaccinated individuals against strain $1$, $V$; infected individuals with strain $i$ but still susceptible to strain $j$, $I_i$, for $i,j=1,2$ and $i\neq j$; immune individuals to strain $j$ and with temporary immunity to strain $i$, $C_i$, for $i,j=1,2$ and $i\neq j$; immune individuals to strain $1$ and still susceptible to strain $2$, vaccinated and unvaccinated, $R_{v1}$ and $R_1$, respectively; immune individuals to strain $2$ and still susceptible to strain $1$, $R_2$; immune individuals to strain $j$ and infected with strain $i$, $Y_i$, for $i,j=1,2$ and $i\neq j$; and immune individuals to all the strain, $R_{12}$. The total population infected by strain $i$ is denoted $J_i$:
\begin{equation}\nonumber
J_i=I_i+Y_i, \quad i=1,2.
\end{equation} 

Thus, the total population is given by: \begin{equation}\nonumber
N(t)= S(t) + V(t) + J_1(t) + J_2(t) + C_1(t) + C_2(t) + R_1(t) + R_2(t) + R_{v1}(t) + R_{12}(t).
\end{equation}   

\begin{figure}[hbt!]
\centering
\includegraphics[scale=0.6]{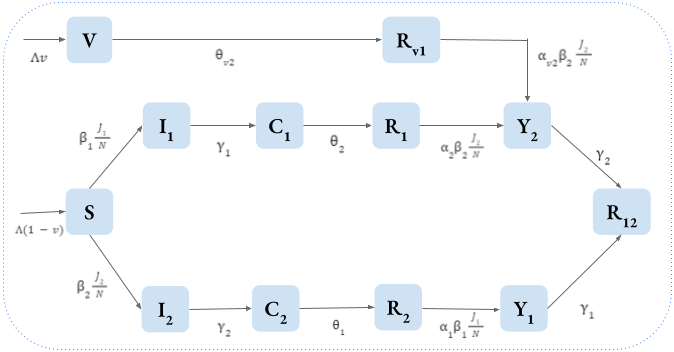}
\caption{Schematic representation of the infection status due to the concomitant transmission of viruses $1$ and $2$, considering that the population is vaccinated against the virus $1$.}
\label{fig:diagrama}
\end{figure}

The flowchart of the model can be seen in Figure \ref{fig:diagrama}.

The population is born and dies at a constant rate of $\Lambda$ and $\mu$, respectively. Part of the population is vaccinated against the virus infection $1$ at birth, at a per capita rate $v$, $0< v\leq 1$. The remaining unvaccinated susceptible population becomes infected by the virus $i$ at a per capita rate $\beta_iJ_i/N$. Infected individuals with virus $i$  recover at a rate of $\gamma_i$. Individuals who recover from infection with virus $i$ become immune to this virus and have temporary immunity to virus $j$, for $i, j=1,2$ and $i\neq j$. This cross-immunity against the virus $i$ wanes at a per capita rate $\theta_i$. The vaccine's immune response is assumed to also confer ttemporary immunity to the virus $2$, and this cross-immunity wanes at the rate per capita $\theta_{v2}$. We assume that after loss of cross-immunity to virus $j$, individuals who are immune to the virus $i$ may be more or less susceptible to secondary infection by virus $j$ due to antibody-dependent enhancement. Thus, the unvaccinated individuals are infected at a per capita rate $\alpha_{j}\beta_{j}J_{j}/N$, for $j=1,2$, while the vaccinated individuals are infected with the virus $2$ at a per capita rate $\alpha_{v2}\beta_{2}J_{2}/N$. The parameters $\alpha_{k}$, for $k=1,2,v2$ represent the fraction that decreases ($0<\alpha_k<1$) or increases ($\alpha_k>1$) the susceptibility to secondary infections. If there is no effect from the antibodies, then $\alpha_k=1$. After recovery from both infections, individuals can no longer become infected. Table \ref{tab:parameters} summarizes the parameters and compartments of the model.

\begin{table}[ht]
\centering
\caption{Parameters and compartments of the model.}
\small{
\begin{tabular}{c|lr}
\textbf{Parameter} & \textbf{Description (for $i, j = 1, 2$)}\\
\hline
$\Lambda$ & Birth rate \\
$\mu$ & Per capita death rate \\
$\beta_i$ & Transmission rate of virus $i$\\
$\gamma_i$ & Per capita recovery rate of infected people with virus $i$\\
$\theta_{i}$ & Per capita loss rate of cross-immunity to virus $i$ after previous infection with virus $j$ \\
$\theta_{v2}$ & Per capita loss rate of cross-immunity to virus $2$ obtained by vaccination\\
$\alpha_i$ & ADE factor that can alter the susceptibility of unvaccinated individuals to the virus $i$ \\
$\alpha_{v2}$ & ADE factor that can alter the susceptibility of vaccinated individuals to virus $2$\\
$v$ & Per capita vaccination rate\\
\textbf{Compartments} & \textbf{Description}\\
\hline
$S$ & Susceptible individuals to both virus\\
$V$ & Vaccinated individuals against the virus $1$\\
$I_i$ & Individuals with primary infection by the virus $i$ \\
$C_i$ & Individuals recovered from infection with virus $i$ and have cross-immunity to virus $j$ \\
$R_i$ & Unvaccinated individuals immune to virus $i$ and susceptible to virus $j$ \\
$R_{v1}$ & Vaccinated individuals to virus $1$ and susceptible to virus $2$ \\
$Y_i$ & Individuals infected by virus $i$ and immune to virus $j$\\
$R_{12}$ & Individuals immune to both virus\\
\hline
\end{tabular}
}
\label{tab:parameters}
\end{table}

The following equations, with appropriate initial conditions, represent the disease dynamics model:

\begin{align}
\dfrac{dS}{dt}&= (1-v)\Lambda - \beta_1J_1\frac{S}{N}-\beta_2J_2\frac{S}{N} -\mu S \nonumber\\
\dfrac{dV}{dt}&= v\Lambda - (\theta_{v2}+\mu)V\nonumber \\
\dfrac{dI_1}{dt}&=\beta_1J_1\frac{S}{N}-(\gamma_1+\mu)I_1\nonumber  \\
\dfrac{dI_2}{dt}&=\beta_2J_2\frac{S}{N}-(\gamma_2+\mu)I_2 \nonumber \\
\dfrac{dC_1}{dt}&=\gamma_1I_1-(\theta_2+\mu)C_1 \nonumber \\
\dfrac{dC_2}{dt}&=\gamma_2I_2 -(\theta_1+\mu)C_2  \nonumber \\
\dfrac{dR_1}{dt}&=\theta_{2}C_1 -\alpha_2 \beta_2J_2\frac{R_1}{N}-\mu R_1 \nonumber \\
\dfrac{dR_2}{dt}&=\theta_{1}C_2 -\alpha_1 \beta_1J_1\frac{R_2}{N}-\mu R_2 \nonumber \\
\dfrac{dR_{v1}}{dt}&=\theta_{v2}V-\alpha_{v2} \beta_2J_2\frac{R_{v1}}{N}-\mu R_{v1} \nonumber\\
\dfrac{dY_1}{dt}&=\alpha_1\beta_1J_1\frac{R_2}{N}-(\gamma_1+\mu)Y_1\nonumber \\
\dfrac{dY_2}{dt}&=\alpha_2\beta_2J_2\frac{R_1}{N}+\alpha_{v2}\beta_2J_2\frac{R_{v1}}{N}-(\gamma_2+\mu)Y_2 \nonumber\\
\dfrac{dR_{12}}{dt}&= \gamma_1Y_1+\gamma_2Y_2-\mu R_{12}, \label{eq:equations}
\end{align}

It follows from the equations that $$\frac{dN(t)}{dt}=\Lambda-\mu N(t).$$ Therefore, \[\lim_{t\rightarrow +\infty}N(t)=\frac{\Lambda}{\mu}.\] Then, without loss of generality, we assume that $N(t)=\Lambda/\mu$, for $t\geq 0$.

Since the system variables represent populations, it is necessary that their values are non-negative and that the system solution is bounded. Proposition \ref{prop:well-defined} shows the limitation and positivity of solutions.

\begin{proposition}\label{prop:well-defined}
Consider the system of equations \eqref{eq:equations}. Given an initial condition in $\mathbb{R}^{12}_+$, then the following conditions hold:
\begin{itemize}
    \item[1.]  There exist a unique bounded solution in $\mathbb{R}^{12}_+$ for the system \eqref{eq:equations}, for all $t \geq 0$;
    \item[2.] $\mathbb{R}^{12}_+$ is positively invariant under the flow of \eqref{eq:equations};
    \item[3.] If S(0) is strictly positive, then  $S(t)$, $V(t)$ and $R_{v1}(t)$ are strictly positive for all $t>0$.
\end{itemize}
\end{proposition}
The proof of Proposition \ref{prop:well-defined} can be found in Appendix \ref{ap:well-defined}.

In the remaining of the work, we consider $S(0)>0$. It follows from the previous Proposition that the model is well-posedness in the set
\begin{eqnarray}
    \Gamma&=&\left\{(S,V,I_1,I_2,C_1,C_2,R_1,R_2,R_{v1},Y_1,Y_2,R_{12})\in \mathbb{R}_+^{12};\right. \nonumber \\
    && \left. S+V+I_1+I_2+C_1+C_2+R_1+R_2+R_{v1}+Y_1+Y_2+R_{12}=\Lambda / \mu \right\}.\nonumber
\end{eqnarray}

\section{Relevant equilibria and reproduction numbers}\label{sec:r0}

In this section, we will find the relevant equilibria of the system from an epidemiological point of view, and calculate the basic and invasion reproduction numbers, which are threshold parameters for the stability of the equilibria. 

\subsection{Disease-free equilibrium and the basic reproductive number}

The disease-free equilibrium (DFE), $E^0$, is the equilibrium point where there is no infections in the population, that is, when $I_1=I_2=Y_1=Y_2=0$. Thus, $E^0 = (S^0,V^0,0,0,C_1^0,C_2^0,R_1^0,R_2^0,R_{v1}^0,0,0,R_{12}^0)$
has coordinates
\begin{equation}\label{eq:E0}
 S^0=(1-v)\frac{\Lambda}{\mu} \textrm{,  } V^0=\frac{v\mu}{\theta_{v2}+\mu}\frac{\Lambda}{\mu} \textrm{,  } R_{v1}^0=\frac{\theta_{v2}v}{\theta_{v2}+\mu}\frac{\Lambda}{\mu}   
\end{equation}
and $C^0_1=C_2^0=R_1^0=R_2^0=R_{12}^0=0$. 

The basic reproduction number, $\mathcal{R}_0$, is defined as the average number of secondary infections produced when a infectious individual is introduced into a fully susceptible population. Its importance lies in the fact that it is a threshold parameter for the stability of disease-free equilibrium. In the following, applying the next generation matrix method \cite{r0}, we will define the basic reproduction number for the system \eqref{eq:equations}. 

The vector referring to the compartments with infected individuals, $x=(J_1,J_2),$ satisfies $$\dot{x}=f(x)-v(x),$$ where $f$ represents the rate of new infections and $v$ represents the transfer rate of individuals by other means:
\begin{equation} \label{matriz:fv}
f=\left(\begin{array}{c}
\frac{\beta_1J_1S}{N}+
\frac{\alpha_1\beta_1J_1R_2}{N}\\
\frac{\beta_2J_2S}{N}+ \frac{\beta_2J_2(\alpha_2R_1+\alpha_{v2}R_{v1})}{N}
\end{array} \right)
\textrm{ and }
v=\left(\begin{array}{c}
(\gamma_1+\mu)J_1 \\
(\gamma_2+\mu)J_2
\end{array} \right).\nonumber
\end{equation}

The matrices $F$ and $V$ are the Jacobian matrices of $f(x)$ and $v(x)$, respectively, evaluated in the $E^0$:
\begin{equation} \label{matriz:FV}
F=\left(\begin{array}{cc}
\frac{\beta_1S^0}{N}&0 \\
0&\frac{\beta_2S^0}{N}+\frac{\beta_2\alpha_{v2}R_{v1}^0}{N}
\end{array} \right)
\textrm{ and }
V=\left(\begin{array}{cc}
\gamma_1+\mu&0 \\
0&\gamma_2+\mu
\end{array} \right).\nonumber
\end{equation}

We define the basic reproductive number as the spectral radius of the next generation matrix $FV^{-1}$:
\begin{equation}\label{eq:R0}
    \mathcal{R}_0=\rho(FV^{-1}) = \max\left\{\mathcal{R}_1,\mathcal{R}_2\right\},
\end{equation}
where
\begin{equation}\label{eq:R1}
    \mathcal{R}_1=\frac{\beta_1S^0}{N(\gamma_1+\mu)}=\frac{\beta_1}{\gamma_1+\mu}(1-v)
\end{equation}
and
\begin{equation}\label{eq:R2}
    \mathcal{R}_2=\frac{\beta_2(S^0+\alpha_{v2}R_{v1}^0)}{N(\gamma_2+\mu)}=\frac{\beta_2}{\gamma_2+\mu}\left[1+v\left(\frac{\alpha_{v2}\theta_{v2}}{\theta_{v2}+\mu}-1\right)\right]. 
\end{equation}

The expression \eqref{eq:R1} for $\mathcal{R}_1$ is given by the product of the transmissibility of the strain $1$, $\beta_1$, the average time an individual spends in the infectious compartment, $1/(\gamma_1+\mu)$, and the fraction of susceptible individuals to this strain (unvaccinated) in the disease-free equilibrium, $S^0/N$. Thus, $\mathcal{R}_1$ represents the average number of new infections caused by an infected individual by the strain $1$, in his infectious period, when there is no other infectious individual in the population. 

The expression \eqref{eq:R2} for $\mathcal{R}_2$ is given by the sum of two components. The first of them is the product of the transmissibility of the strain $2$, $\beta_2$, the average time an individual spends in the infectious compartment, $1/(\gamma_2+\mu)$, and the fraction of unvaccinated susceptible individuals in the disease-free equilibrium, $S^0/N$. The second term is the product of the transmissibility of the strain $2$, $\beta_2$, the average time an individual spends in the infectious compartment, $1/(\gamma_2+\mu)$, the fraction of vaccinated susceptible individuals in the disease-free equilibrium, $R_{v1}^0/N$, and the factor of increase or not of susceptibility $\alpha_{v2}$. Just like $\mathcal{R}_1$, the value $\mathcal{R}_2$ represents the average number of new infections caused by an infected individual by the strain $2$, in his infectious period, when there is no other infectious individual in the population. 

\begin{remark}
In the model without vaccination ($v=0$), the basic reproductive number is the maximum between the reproductive numbers of each strain, 
\begin{equation}
\mathcal{R}_1^{wv}=\dfrac{\beta_1}{\gamma_1+\mu} \textrm{ and } \mathcal{R}_2^{wv}=\dfrac{\beta_2}{\gamma_2+\mu}. \label{eq:R0-wv}
\end{equation}
\end{remark}

\begin{remark}
The vaccination decreases the value of $\mathcal{R}_1^{wv}$, reducing the number of new infections by the strain $1$. The effect of the vaccination over strain $2$ depends on the parameters of the vaccine, $\alpha_{v2}$ and $\theta_{v2}$, referents to ADE and loss of cross-immunity against strain $2$.
\end{remark}

\subsection{Endemic boundary equilibria}

In addition to disease-free equilibrium, we look for more two relevant equilibriums on the boundary: the endemic equilibrium where there are only infections by the strain $1$, $E^1$, and the endemic equilibrium where there are only infections by the strain $2$, $E^2$.

At the equilibrium $E^1$, the values of $I_2,C_2,R_2,Y_1,Y_2$ and $R_{12}$ are zero. Then
\begin{equation}\label{eq:E1}
    E^1=(S^*,V^*,I_1^*,0,C_1^*,0,R_1^*,0,R_{v1}^*,0,0,0),
\end{equation}
where
\begin{eqnarray}
    &&S^*=\frac{(\gamma_1+\mu)\Lambda}{\beta_1\mu}, \quad V^*=\frac{v\Lambda}{\theta_{v2}+\mu}, \quad I_1^*=\frac{(1-v)\Lambda}{\gamma_1+\mu}\left(1-\frac{1}{\mathcal{R}_1}\right),\nonumber \\
    &&C_1^*=\frac{\gamma_1}{\theta_2+\mu}I_1^*, \quad R_1^*=\frac{\theta_2}{\mu}C_1^* \quad \textrm{and} \quad R_{v1}^*=\frac{\theta_{v2}}{\mu}V^*.\nonumber
\end{eqnarray}
The expression of $\mathcal{R}_1$ is given in \eqref{eq:R1}. Note that the endemic equilibrium $E^1$ exists if and only if $\mathcal{R}_1>1$.

At the equilibrium $E^2$, the values of $I_1,C_1,R_1$ and $Y_1$ are zero. Then
\begin{equation}\label{eq:E2}
 E^2=(S^*,V^*,0,I_2^*,0,C_2^*,0,R_2^*,R_{v1}^*,0,Y_2^*,R_{12}^*),
\end{equation}
where
\begin{eqnarray}
    &&S^*=\frac{(1-v)\Lambda}{x+\mu}, \quad V^*=\frac{v\Lambda}{\theta_{v2}+\mu}, \quad I_2^*=\frac{(1-v)x\Lambda}{(x+\mu)(\gamma_2+\mu)}, \quad C_2^*=\frac{(1-v)x\gamma_2\Lambda}{(x+\mu)(\gamma_2+\mu)(\theta_1+\mu)}, \nonumber\\ &&R_2^*=\frac{(1-v)x\gamma_2\theta_1\Lambda}{(x+\mu)(\gamma_2+\mu)(\theta_1+\mu)\mu}, \quad R_{v1}^*=\frac{v\theta_{v2}\Lambda}{(\theta_{v2}+\mu)(\alpha_{v2}x+\mu)},\nonumber\\ &&Y_2^*=\frac{v\alpha_{v2}x\theta_{v2}\Lambda}{(\alpha_{v2}x+\mu)(\theta_{v2}+\mu)(\gamma_2+\mu)}, \quad R_{12}^*=\frac{v\alpha_{v2}x\theta_{v2}\gamma_2\Lambda}{(\alpha_{v2}x+\mu)(\theta_{v2}+\mu)(\gamma_2+\mu)\mu},\nonumber
\end{eqnarray}
\begin{equation}
    x=\frac{\beta_2\mu(I_2^*+Y_2^*)}{\Lambda} \nonumber
\end{equation}
and $x$ is solution of the quadratic equation
\begin{equation}\label{eq:polinomio}
    ax^2+bx+c=0,
\end{equation}
with coefficients $a,b$ and $c$ given by
\begin{eqnarray}
    a&=&\alpha_{v2}\nonumber\\
    b&=&\mu\alpha_{v2}\left[1-\frac{\beta_2(1-v)}{\gamma_2+\mu}\right]+\mu\left[1-\frac{\beta_2\alpha_{v2}\theta_{v2}v}{(\gamma_2+\mu)(\theta_{v2}+\mu)}\right]\nonumber\\
    c&=&\mu^2\left(1-\mathcal{R}_2\right).\nonumber
\end{eqnarray}

If $\mathcal{R}_2\leq 1$, the fractions in the expression of $b$ must be smaller than one or equal to one, and it is not possible for both to be one. Therefore, $b>0$. We also have $c\geq 0$. Since that $a>0$, the equation \eqref{eq:polinomio} does not have roots with positive real parts. This implies that there is no endemic equilibrium like $E^2$. Thus, for an equilibrium $E^2$ to exist, we must have $\mathcal{R}_2>1$. In this case, $c<0$. Since the coefficient $a$ is positive, the equation \eqref{eq:polinomio} has two real roots and only one of them is positive. In resume, if $\mathcal{R}_2>1$, there is a unique endemic equilibrium where there are infections only by the strain $2$. The value of $I_2^*+Y_2^*$ at the equilibrium is calculated through the positive solution of the equation \eqref{eq:polinomio}.

The results above give us the following Theorem.

\begin{theorem}
Let $\mathcal{R}_1$ and $\mathcal{R}_2$ be given in \eqref{eq:R1} and \eqref{eq:R2}, respectively. The system \eqref{eq:equations} has an endemic equilibrium with infections caused only by the strain $1$ if and only if $\mathcal{R}_1>1$. The system \eqref{eq:equations} has an endemic equilibrium with infections caused only by the strain $2$ if and only if $\mathcal{R}_2>1$. In both cases the equilibria are unique.
\end{theorem}

\subsection{Invasion reproduction numbers}
Like the basic reproduction number, the invasion reproduction number is a relevant threshold parameter for the analysis of equilibrium stability. It means the average number of new infections caused by an individual infected by a strain during his infectious period, in a population that is susceptible to this strain, but is at the endemic equilibrium of another strain. This concept is explained, for example, in \cite{invasion,r0}. The invasion numbers, as the basic reproduction number, were calculated using the next generation matrix \cite{r0}.

To define the invasion reproduction number of strain $2$ at the equilibrium of strain $1$, $\mathcal{R}^2_1$, we consider $x=J_2$ and proceed building the matrices $f(x)$ and $v(x)$ (which are scalars) of the new infections (caused by the strain $2$) and the remaining transfer terms, respectively. The next generation matrix is the matrix $F_2V_2^{-1}$, where $F_2$ and $V_2$ are the Jacobian of the matrices $f$ and $v$, evaluated at the equilibrium $E^1$. Thus,
\begin{equation}\label{eq:IRN-2}
    \mathcal{R}^2_1=\rho(F_2V_2^{-1})=\frac{\beta_2S^*}{(\gamma_2+\mu)N}+\frac{\beta_2(\alpha_2R_1^*+\alpha_{v2}R_{v1}^*)}{(\gamma_2+\mu)N},
\end{equation}
where $S^*$, $R_1^*$ and $R_{v1}^*$ are given in the expression of $E^1$ \eqref{eq:E1}.

For interpretation of the $\mathcal{R}_1^2$, remember that at the equilibrium $E^1$, an individual infected by strain $2$ can infect the unvaccinated susceptible individuals (susceptible to all strains), $S^*$, and the individuals immune to strain $1$ and susceptible to strain $2$. In the last case we have two options: the individuals that had an infection by the strain $1$, recovered and, after of a period, lost the cross-immunity against strain $2$, $R_1^*$; or the individuals that received the vaccine and, after of a period, lost the cross-immunity offered by the vaccine against strain $2$, $R_{v1}^*$. The parameter $\beta_2$ is the transmissibility of the strain $2$, and $1/(\gamma_2+\mu)$ is the duration of the infectious period of an individual infected by the strain $2$. The parameters $\alpha_2$ and $\alpha_{v2}$ are the factors of ADE that can appear after recuperation from an infection by the strain $1$ or after a vaccination, respectively.

Analogously, we calculated the invasion reproduction number of the strain $1$ at the equilibrium of the strain $2$:
\begin{equation}\label{eq:IRN-1}
    \mathcal{R}_2^1=\rho(F_1V_1^{-1})=\frac{\beta_1S^*}{(\gamma_1+\mu)N}+\frac{\alpha_1\beta_1R_2^*}{(\gamma_1+\mu)N},
\end{equation}
where $S^*$ and $R_2^*$ are given in the expression of $E^2$ \eqref{eq:E2}.

At the equilibrium $E^2$, an individual infected by strain $1$ can infect the unvaccinated susceptible individuals (susceptible to all strains), $S^*$, and the individuals immune to strain $2$ and susceptible to strain $1$, $R_2^*$. The last, had an infection by strain $2$, recovered and, after of a period, lost the cross-immunity against strain $1$. The parameter $\beta_1$ is the transmissibility of the strain $1$, and $1/(\gamma_1+\mu)$ is the duration of the infectious period of an individual infected by the strain $1$. The parameter $\alpha_1$ is the factor of ADE that can appear after recuperation from an infection by the strain $2$.  

\section{Stability analysis}\label{sec:estabilidade}
In this section, we will give results about the stability of the disease-free and endemic equilibria. Before starting this analysis, we will comment about the stability of the DFE in the model without vaccination.

\subsection{The DFE in the model without vaccination}
The model without vaccination \cite{maira,vanessa} describes the dynamics of population with the circulation of two strains of a virus. In this model $v=0$ and the states $V$ and $R_{v1}$ are not considered. The model is well-posedness in the set
\begin{eqnarray}
    \Gamma^{wv}&=&\left\{(S,I_1,I_2,C_1,C_2,R_1,R_2,Y_1,Y_2,R_{12})\in \mathbb{R}_+^{10};\right. \nonumber \\
    && \left. S+I_1+I_2+C_1+C_2+R_1+R_2+Y_1+Y_2+R_{12}=\Lambda / \mu \right\}.\nonumber
\end{eqnarray}
The DFE is the point $$E^{wv}_0=\left(\frac{\Lambda}{\mu},0,0,0,0,0,0,0,0,0\right).$$

\noindent In \cite{vanessa}, we can find the following Theorem.
\begin{theorem}
Let $\mathcal{R}_1^{wv}$ and $\mathcal{R}_2^{wv}$ be as given in \eqref{eq:R0-wv}. If $\mathcal{R}_0^{wv}=\max\{\mathcal{R}_1^{wv},\mathcal{R}_2^{wv}\}<1$, then the DFE, $E^{wv}_0$, is locally asymptotically stable. If $\mathcal{R}_0^{wv}>1$, then the DFE is unstable.
\end{theorem}
\begin{proof}
(Sketch) The proof is obtained linearizing the system and evaluating at $E^{wv}_0$. In the Jacobian matrix, eight eigenvalues are negative and the others two have the signal determined by $\mathcal{R}_1^{wv}$ and $\mathcal{R}_2^{wv}$. If $\mathcal{R}_0^{wv}<1$, the eigenvalues are all negative. If $\mathcal{R}_0^{wv}>1$, at least one of them is positive.
\end{proof}

Next, we will analyze the dynamics of the model with vaccination, given by the system of equations \eqref{eq:equations}.

\subsection{Local stability}

With the definition of $\mathcal{R}_0$, we proved the first result about the local stability of the disease-free equilibrium. 

\begin{theorem}\label{theo:ro-local}
Let $\mathcal{R}_0$ be as defined in \eqref{eq:R0}. The disease-free equilibrium of the model \eqref{eq:equations}, $E^0$, is locally asymptotically stable if $\mathcal{R}_0<1$, and unstable if $\mathcal{R}_0>1$.
\end{theorem}
\begin{proof}
Linearizing the system \eqref{eq:equations} at the equilibrium point $E^0$ and calculating the characteristic polynomial, we obtained:
\begin{eqnarray}
    P(\lambda)&=&(\lambda+\theta_{v2}+\mu)(\lambda+\theta_2+\mu)(\lambda+\theta_1+\mu)(\lambda+\mu)^5(\lambda+\gamma_1+\mu)(\lambda+\gamma_2+\mu)\nonumber\\
    &&[(\gamma_1+\mu)(\mathcal{R}_1-1)-\lambda][(\gamma_2+\mu)(\mathcal{R}_2-1)-\lambda]\nonumber
\end{eqnarray}
Here, if $\mathcal{R}_0=\max\{\mathcal{R}_1,\mathcal{R}_2\}<1$, then all the eigenvalues are negative real numbers and the equilibrium is locally asymptotically stable. If $\mathcal{R}_0>1$, at least one of the eigenvalues is positive and the equilibrium is unstable.
\end{proof}

With the definition of the invasion reproduction numbers, we proved the following results about the local stability of the endemic boundary equilibria.

\begin{theorem}\label{teo:rinv-21}
Let $\mathcal{R}_1$ and $\mathcal{R}^2_1$ be as defined in \eqref{eq:R1} and \eqref{eq:IRN-2}, respectively. Suppose $\mathcal{R}_1>1$. The endemic boundary equilibrium of the model \eqref{eq:equations}, $E^1$, is locally asymptotically stable if $\mathcal{R}^2_1<1$, and unstable if $\mathcal{R}^2_1>1$.
\end{theorem}
\begin{proof}
Linearizing the system \eqref{eq:equations} at the equilibrium point $E^1$ and calculating the characteristic polynomial, we obtained:
\begin{eqnarray}
    P(\lambda)&=&(-\mu-\lambda)^3[-(\theta_{v2}+\mu)-\lambda][-(\theta_2+\mu)-\lambda][-(\theta_1+\mu)-\lambda][-(\gamma_1+\mu)-\lambda][-(\gamma_2+\mu)-\lambda]\nonumber\\
    &&\left[-\left(\frac{\alpha_1\beta_1I_1^*}{N}+\mu\right)-\lambda\right][(\gamma_2+\mu)(\mathcal{R}_1^2-1)-\lambda]Q(\lambda),\nonumber
\end{eqnarray}
where 
\begin{equation}
    Q(\lambda)=\lambda^2+b\lambda+c,\nonumber
\end{equation}
and the coefficients $b$ and $c$ are 
\begin{eqnarray}
    &&b=(\gamma_1+2\mu)+\frac{\beta_1I_1^*}{N}-\frac{\beta_1S^*}{N},\nonumber\\
    &&c=\mu(\gamma_1+\mu)+\frac{\beta_1I_1^*(\gamma_1+\mu)}{N}-\frac{\beta_1S^*\mu}{N}.\nonumber
\end{eqnarray}

Substituting the expressions of the $S^*$ and $I_1^*$, given in \eqref{eq:E1}, in the expressions for $b$ and $c$, we have 
\begin{eqnarray}
    b&=&\mu+\frac{\beta_1\mu(1-v)}{\gamma_1+\mu}\left(1-\frac{1}{\mathcal{R}_1}\right),\nonumber\\
    c&=&\beta_1\mu(1-v)\left(1-\frac{1}{\mathcal{R}_1}\right).\nonumber
\end{eqnarray}
As $\mathcal{R}_1>1$, then $b>0$ and $c>0$. Therefore, the two roots of the polynomial $Q(\lambda)$ have negative real parts. It follows that if $\mathcal{R}^2_1<1$, then all roots of $P(\lambda)$ have negative real parts and the point $E^1$ is locally asymptotically stable, while if $\mathcal{R}^2_1>1$, $P(\lambda)$ has one positive real root and the point $E^1$ is unstable. 
\end{proof}

\begin{theorem}\label{teo:rinv-12}
Let $\mathcal{R}_2$ and $\mathcal{R}^1_2$ be as defined in \eqref{eq:R2} and \eqref{eq:IRN-1}, respectively. Suppose $\mathcal{R}_2>1$. The endemic boundary equilibrium of the model \eqref{eq:equations}, $E^2$, is locally asymptotically stable if $\mathcal{R}^1_2<1$, and unstable if $\mathcal{R}^1_2>1$.
\end{theorem}

\begin{proof}
To simplify the calculations, we will consider the system \eqref{eq:equations} with the variables $$J_1,Y_1,J_2,Y_2,S,V,C_1,C_2,R_1,R_2,R_{v1},R_{12}.$$ 

Linearizing the system at the equilibrium $E^2$, we have the characteristic polynomial
\begin{eqnarray}
    P(\lambda)&=&-(-\mu-\lambda)^2[-(\theta_{v2}+\mu)-\lambda][-(\theta_2+\mu)-\lambda][-(\theta_1+\mu)-\lambda][-(\gamma_1+\mu)-\lambda][-(\gamma_2+\mu)-\lambda]\nonumber\\
    &&\left[-\left(\frac{\alpha_2\beta_2J_2^*}{N}+\mu\right)-\lambda\right][(\gamma_1+\mu)(\mathcal{R}_2^1-1)-\lambda]Q(\lambda).\nonumber
\end{eqnarray}

Using that $$\beta_2S^*/N+\alpha_{v2}\beta_2R_{v1}^*/N-(\gamma_2+\mu)=0,$$ $Q(\lambda)$ is the polynomial $Q(\lambda)=\lambda^3+b\lambda^2+c\lambda+d$, with positive coefficients given in Appendix \ref{ap:coef}. The signals of the real parts of roots of $Q$ can be studied by the Routh-Hurwitz criterion \cite{routh}. The table of the method is
\begin{equation} 
\left(\begin{array}{cccc}
1&c&0&0\\
b&d&0&0\\
\frac{bc-d}{b}&0&0&0\\
d&0&0&0
\end{array} \right).\nonumber
\end{equation}
More calculations shows that the first column is positive. As there is no change of signal in this column, by the Routh criterion, the real parts of the roots of $Q$ are negative.  

It follows that the equilibrium $E^2$ is stable if $\mathcal{R}_2^1<1$, and unstable if $\mathcal{R}_2^1>1$.
\end{proof}

\begin{remark}\label{rem:alfa-rinv}
    Note that in the case $\alpha_1\leq 1$, it is valid that $\mathcal{R}_2^1\leq \mathcal{R}_1$ (see Appendix \ref{ap:alfa-rinv}). That is, if $\mathcal{R}_1<1$, the strain $1$ can not invade the endemic equilibrium of the strain $2$. If $\alpha_1>1$, even with $\mathcal{R}_1<1$, the strain $1$ may or may not persist. The analogous is valid for the case $\alpha_2\leq 1 $ and $\alpha_2>1$.
\end{remark}

\subsection{Analysis of the subsystems}
In this section, we will study the dynamics of two following subsystems, where there are infections by only one of the strains:

\begin{eqnarray}\label{eq:equations-only1}
    \frac{dS}{dt}&=&(1-v)\Lambda-\frac{\beta_1I_1S}{N}-\mu S\nonumber\\
    \frac{dV}{dt}&=&v\Lambda -(\theta_{v2}+\mu)V\nonumber\\
    \frac{dI_1}{dt}&=&\frac{\beta_1I_1S}{N}-(\gamma_1+\mu)I_1\nonumber\\
    \frac{dC_1}{dt}&=&\gamma_1I_1-(\theta_2+\mu)C_1\nonumber\\
    \frac{dR_1}{dt}&=&\theta_2C_1-\mu R_1\nonumber\\
    \frac{dR_{v1}}{dt}&=&\theta_{v2}V-\mu R_{v1}
\end{eqnarray}
and
\begin{eqnarray}\label{eq:equations-only2}
    \frac{dS}{dt}&=&(1-v)\Lambda-\frac{\beta_2J_2S}{N}-\mu S\nonumber\\
    \frac{dV}{dt}&=&v\Lambda-(\theta_{v2}+\mu)V\nonumber\\
    \frac{dI_2}{dt}&=&\frac{\beta_2J_2S}{N}-(\gamma_2+\mu)I_2\nonumber\\
    \frac{dC_2}{dt}&=&\gamma_2I_2-(\theta_1+\mu)C_2\nonumber\\
    \frac{dR_2}{dt}&=&\theta_1C_2-\mu R_2\nonumber\\
    \frac{dR_{v1}}{dt}&=&\theta_{v2}V-\frac{\alpha_{v2}\beta_2J_2R_{v1}}{N}-\mu R_{v1}\nonumber\\
    \frac{dY_2}{dt}&=&\frac{\alpha_{v2}\beta_2J_2R_{v1}}{N}-(\gamma_2+\mu)Y_2\nonumber\\
    \frac{dR_{12}}{dt}&=&\gamma_2Y_2-\mu R_{12},
\end{eqnarray}
defined in the sets 
\begin{equation}
    \Gamma_1=\left\{(S,V,I_1,C_1,R_1,R_{v1})\in \mathbb{R}_+^{6};S+V+I_1+C_1+R_1+R_{v1}=\frac{\Lambda}{\mu}\right\}\nonumber
\end{equation}
and
\begin{equation}
    \Gamma_2=\left\{(S,V,I_2,C_2,R_2,R_{v1},Y_2,R_{12})\in \mathbb{R}_+^{8};S+V+I_2+C_2+R_2+R_{v1}+Y_2+R_{12}=\frac{\Lambda}{\mu}\right\},\nonumber
\end{equation}
respectively. The dynamics of the systems \eqref{eq:equations-only1} and \eqref{eq:equations-only2} are the dynamics of the full system \eqref{eq:equations} when the strain $2$ is extinct and when the strain $1$ is extinct, respectively. From the Proposition \eqref{prop:well-defined}, the systems \eqref{eq:equations-only1} and \eqref{eq:equations-only2} are well-defined in the sets $\Gamma_1$ and $\Gamma_2$, respectively. The disease-free equilibria $E^0_1$ and $E^0_2$, of the subsystems \eqref{eq:equations-only1} and \eqref{eq:equations-only2}, respectively, correspond to the disease-free equilibrium of the full system, $E^0$. That is, they have coordinates $S$, $V$ and $R_{v1}$ equal to $S^0$, $V^0$ and $R_{v1}^0$, respectively, given in \eqref{eq:E0}, and the others coordinates are null. The interior equilibrium of each system can also be directly deduced from the endemic boundary equilibria of the full system. The interior equilibrium of the system \eqref{eq:equations-only1} is $E^{1}_1=(S^*,V^*,I_1^*,C_1^*,R_1^*,R_{v1}^*)$, where the coordinates of $E^{1}_1$ are the positive coordinates given in \eqref{eq:E1} for $E^1$. The interior equilibrium of the system \eqref{eq:equations-only2} is $E^{2}_2=(S^*,V^*,I_2^*,C_2^*,R_2^*,R_{v1}^*,Y_2^*,R_{12}^*)$. On the same way, the coordinates of $E^{2}_2$ are the positive coordinates given in \eqref{eq:E2} for $E^2$. 

Next, we prove the global stability of the disease-free and endemic equilibria in the subsystems \eqref{eq:equations-only1} and \eqref{eq:equations-only2}. A version (see \cite{chemostat}, Chap. 2, page 29) of the LaSalle's Invariance Principle \cite{lasalle} is the main tool used in the proofs. During the process, Lyapunov functions are constructed using combinations of the classical Lyapunov function $L=x-x^*\ln x$, used since $1980$'s in ecological models \cite{exemplo}, quadratic functions \cite{stability-sis} and the methods described in \cite{stability}.

\begin{theorem}\label{theo:subsystem-dfe-1}
If $\mathcal{R}_1\leq 1$, then the disease-free equilibrium, $E^{0}_1$, is globally asymptotically stable for system \eqref{eq:equations-only1} in $\Gamma_1$.
\end{theorem}

\begin{proof}
It is clear that the set $\Gamma_1$ is invariant by the solution of the system. 

At the equilibrium $E^0_1$, it is valid that
\begin{equation}\label{eq:equilibrium-dfe-only1}
    (1-v)\Lambda-\mu S^0=0.
\end{equation} 
    
Let $L$ be the Lyapunov function     
   \begin{equation}
    L(t)=\left(S-S^0-S^0\ln{\frac{S}{S^0}}\right)+I_1
    \end{equation}
in $G=\{(S,V,I_1,C_1,R_1,R_{v1})\in \Gamma_1;S>0\}$.

Differentiating $L$ with respect to $t$, along solutions of \eqref{eq:equations-only1}, and using equation \eqref{eq:equilibrium-dfe-only1}, gives
\begin{eqnarray}
    L'(t)&=&(S-S^0)\left[\frac{(1-v)\Lambda}{S}-\mu-\frac{\beta_1I_1}{N}\right]+\frac{\beta_1I_1S}{N}-(\gamma_1+\mu)I_1\\\nonumber
    &&=(1-v)\Lambda(S-S^0)\left(\frac{1}{S}-\frac{1}{S^0}\right)+\frac{\beta_1I_1S^0}{N}-(\gamma_1+\mu)I_1\nonumber\\
    &&=(1-v)\Lambda\left(2-\frac{S}{S^0}-\frac{S^0}{S}\right)+I_1(\gamma_1+\mu)(\mathcal{R}_1-1).\nonumber
\end{eqnarray}

We have that $2-S/S^0-S^0/S\leq 0$ and the equality is valid only if $S=S^0$. Since that $\mathcal{R}_1\leq 1$, we have $L'(t)\leq 0$ in $G$.

If $\mathcal{R}_1<1$, then $L'(t)=0$ if and only if $I_1=0$ and $S=S^0$. If $\mathcal{R}_1=1$, then $L'(t)=0$ if and only if $S=S^0$. Note that $V$ tends to $V^0$, when $t$ tends to infinity. Also, if $V=V^0$, integrating the equation for $dR_{v1}/dt$, we have that $R_{v1}$ tends to $R_{v1}^0$ when $t$ tends to infinity. Since that $S^0+V^0+R_{v1}^0=\Lambda/\mu$, the largest invariant set by \eqref{eq:equations-only1} contained in $E=\{(S,V,I_1,C_1,R_1,R_{v1})\in G;L'(t)=0\}$ is the singleton $\{E^0_1\}$. Thus, the endemic equilibrium $E^{0}_1$ is globally asymptotically stable in $G$, by LaSalle’s Invariable Principle \cite{chemostat}. All orbit of the system \eqref{eq:equations-only1} starting at a point in $\Gamma_1$, belongs to $G$ for $t>0$. Thus, the equilibrium $E^{0}_1$ is globally asymptotically stable in $\Gamma_1$.
\end{proof}

\begin{theorem}\label{theo:subsystem-dfe-2}
If $\mathcal{R}_2\leq 1$, then the disease-free equilibrium $E^{0}_2$ is globally asymptotically stable for system \eqref{eq:equations-only2} in $\Gamma_2$. 
\end{theorem}

\begin{proof}
In this proof, we will use the method in \cite{stability}. 

At the equilibrium $E^0_2$, it is valid that
\begin{eqnarray}\label{eq:equilibrium-E0-2}
    &&(1-v)\Lambda-\mu S^0=0\nonumber\\
    &&v\Lambda-(\theta_{v2}+\mu)V^0=0\nonumber\\
    &&\theta_{v2}V^0-\mu R_{v1}^0=0.
\end{eqnarray}

Let $L$ be the Lyapunov function     
   \begin{equation}
    L(t)=\left(S-S^0-S^0\ln{\frac{S}{S^0}}\right)+\left(V-V^0-V^0\ln{\frac{V}{V^0}}\right)+\left(R_{v1}-R^0_{v1}-R_{v1}^0\ln{\frac{R_{v1}}{R_{v1}^0}}\right)+I_2+Y_2\nonumber
    \end{equation}
defined in $G=\{(S,V,I_2,C_2,R_2,R_{v1},Y_2,R_{12})\in \Gamma_2;S>0,V>0,R_{v1}>0\}$.

Differentiating $L(t)$, with respect to $t$, along solutions of \eqref{eq:equations-only2}, and using the equations in \eqref{eq:equilibrium-E0-2}, we have

\begin{eqnarray}
    L'(t)&=&(S-S^0)\left[\frac{(1-v)\Lambda}{S}-\frac{\beta_2J_2}{N}-\mu\right]+(V-V^0)\left[\frac{v\Lambda}{V}-(\theta_{v2}+\mu)\right]\nonumber\\
    &&+(R_{v1}-R_{v1}^0)\left(\frac{\theta_{v2}V}{R_{v1}}-\frac{\alpha_{v2}\beta_2J_2}{N}-\mu\right)+\frac{\beta_2J_2S}{N}+\frac{\alpha_{v2}\beta_2J_2R_{v1}}{N}-(\gamma_2+\mu)J_2\nonumber\\
    &=&(1-v)\Lambda(S-S^0)\left(\frac{1}{S}-\frac{1}{S^0}\right)+v\Lambda(V-V^0)\left(\frac{1}{V}-\frac{1}{V^0}\right)+\theta_{v2}(R_{v1}-R_{v1}^0)\left(\frac{V}{R_{v1}}-\frac{V^0}{R_{v1}^0}\right)\nonumber\\
    &&+J_2(\gamma_2+\mu)(\mathcal{R}_2-1) \nonumber\\
    &=&F(S,V,R_{v1})+J_2(\gamma_2+\mu)(\mathcal{R}_2-1),\nonumber
\end{eqnarray}

where $$F(S,V,R_{v1})=(1-v)\Lambda(S-S^0)\left(\frac{1}{S}-\frac{1}{S^0}\right)+v\Lambda(V-V^0)\left(\frac{1}{V}-\frac{1}{V^0}\right)+\theta_{v2}(R_{v1}-R_{v1}^0)\left(\frac{V}{R_{v1}}-\frac{V^0}{R_{v1}^0}\right).$$

We will show that $F(S,V,R_{v1})\leq 0$, and the equality is valid only if $S=S^0$, $V=V^0$ and $R_{v1}=R_{v1}^0$. For this denote $x=\frac{S}{S^0}$, $y=\frac{V}{V^0}$, $z=\frac{R_{v1}}{R_{v1}^0}$. Rewriting $F(S,V,R_{v1}):=F(x,y,z)$, we have
\begin{eqnarray}
    F(x,y,z)&=&(1-v)\Lambda(x-1)\left(\frac{1}{x}-1\right)+v\Lambda(y-1)\left(\frac{1}{y}-1\right)+\theta_{v2}V^0(z-1)\left(\frac{y}{z}-1\right)\nonumber\\
    &=&2(1-v)\Lambda+2v\Lambda+\theta_{v2}V^0-(1-v)\Lambda x-(1-v)\Lambda \frac{1}{x}+(-v\Lambda+\theta_{v2}V^0)y-v\Lambda\frac{1}{y}\nonumber\\
    &&-\theta_{v2}V^0z-\theta_{v2}V^0\frac{y}{z}.\nonumber
\end{eqnarray}

Using the method in \cite{stability}, we rewrite $F(x,y,z)$ as
\begin{equation}
    F(x,y,z)=(1-v)\Lambda\left(2-x-\frac{1}{x}\right)+(v\Lambda-\theta_{v2}V^0)\left(2-y-\frac{1}{y}\right)+\theta_{v2}V^0\left(3-z-\frac{1}{y}-\frac{y}{z}\right).\nonumber
\end{equation}

Lastly, using the two first equations in \eqref{eq:equilibrium-E0-2}, $F(x,y,z)$ can be rewritten as
$$F(x,y,z)=\mu S^0\left(2-x-\frac{1}{x}\right)+\mu V^0\left(2-y-\frac{1}{y}\right)+\mu R^0_{v1}\left(3-z-\frac{1}{y}-\frac{y}{z}\right).$$

Since that the arithmetic average is greater or equal than geometric average, $F(x,y,z)\leq0$ and the equality is valid if and only if $x=y=z=1$.

Thus, since that $\mathcal{R}_2\leq 1$, we have $L'(t)\leq 0$. If $\mathcal{R}_2<1$, then $L'(t)=0$ if and only if $J_2=0$ and $F(S,V,R_{v1})=0$. If $\mathcal{R}_2=1$, then $L'(t)=0$ if and only $F(S,V,R_{v1})=0$. Note that $S=S^0$ for all $t$ implies $J_2=0$. Thus, the largest invariant set of \eqref{eq:equations-only2} contained in 
\begin{eqnarray}
    E&=&\{(S,V,I_2,C_2,R_2,R_{v1},Y_2,R_{12})\in G;L'(t)=0\}\nonumber\\
    &=&\{(S,V,I_2,C_2,R_2,R_{v1},Y_2,R_{12})\in G;S=S^0,V=V^0,R_{v1}=R_{v1}^0\}\nonumber
\end{eqnarray}
is the singleton $\{E^0_2\}$. It follows from the LaSalle's Invariance Principle \cite{chemostat} that the equilibrium $E^0_2$ is globally asymptotically stable in $G$. From the similar calculations to those in Proposition \ref{prop:well-defined}, all orbit of \eqref{eq:equations-only2} belongs to $G$ for all $t>0$. Therefore, $E^0_2$ is globally asymptotically stable in $\Gamma_2$. 
\end{proof}

The following theorems give us information about the stability of the interior equilibrium of each subsystem.

\begin{theorem}\label{theo:stability-E1-subsystem}
Consider $\mathcal{R}_1>1$. The equilibrium $E^{1}_1$ is globally asymptotically stable for system \eqref{eq:equations-only1} in $\{(S,V,I_1,C_1,R_1,R_{v1})\in \Gamma_1;I_1>0\}$.
\end{theorem}
\begin{proof}
We will use the Lyapunov function described in \cite{stability-sis}.

At the equilibrium $E^{1}_1$, it is valid that
\begin{eqnarray}\label{eq:subsystem1-equilibrium}
    &&(1-v)\Lambda-\frac{\beta_1I_1^*S^*}{N}-\mu S^*=0\nonumber\\
    &&\frac{\beta_1I_1^*S^*}{N}-(\gamma_1+\mu)I_1^*=0.
\end{eqnarray} 
    
Let $L$ be the Lyapunov function     
    \begin{equation}\nonumber
        L(t)=\frac{1}{2}\left[(S-S^*)+(I_1-I_1^*)\right]^2+k\left(I_1-I_1^*-I_1^*\ln{\frac{I_1}{I_1^*}}\right),
    \end{equation}
where $k=\dfrac{2\mu+\gamma_1}{\beta_1}\dfrac{\Lambda}{\mu}$, defined in $$G=\{(S,V,I_1,C_1,R_1,R_{v1})\in \Gamma_1;I_1>0\}.$$

Differentiating $L$ with respect to $t$, along solutions of \eqref{eq:equations-only1} gives
\begin{eqnarray}
    L'(t)&=&[(S-S^*)+(I_1-I_1^*)][(1-v)\Lambda-\mu S-(\gamma_1+\mu)I_1]+k(I_1-I_1^*)\left[\frac{\beta_1S}{N}-(\gamma_1+\mu)\right].\nonumber
\end{eqnarray}
Using the equations in \eqref{eq:subsystem1-equilibrium}, we have
\begin{eqnarray}
    L'(t)&=&-[(S-S^*)+(I_1-I_1^*)][(\gamma_1+\mu)(I_1-I_1^*)+\mu(S-S^*)]+\frac{k\beta_1}{N}(I_1-I_1^*)(S-S^*)\nonumber\\
    &=&-\mu(S-S^*)^2-(\gamma_1+\mu)(I_1-I_1^*)^2.\nonumber
\end{eqnarray}

Thus, $L'(t)\leq 0$ and the equality is valid if and only if $S=S^*$ and $I_1=I_1^*$.

Lastly, since that $\{E^{1}_1\}$ is the maximum invariant set of \eqref{eq:equations-only1} contained in $$\{(S,V,I_1,C_1,R_1,R_{v1})\in G;L'(t)=0\}=\{(S,V,I_1,C_1,R_1,R_{v1})\in G;S=S^*,I_1=I_1^*\},$$ by the LaSalle's Invariance Principle \cite{chemostat}, the equilibrium $E^{1}_1$ is globally asymptotically stable in $G$.
\end{proof}

\begin{theorem}\label{theo:stability-E2-subsystem}
Consider $\mathcal{R}_2>1$. The equilibrium $E^{2}_2$ is globally asymptotically stable for system \eqref{eq:equations-only2} in $\{(S,V,I_2,C_2,R_2,R_{v1},Y_2,R_{12})\in \Gamma_2;I_2+Y_2>0\}$. 
\end{theorem}
\begin{proof}
Remember that $J_2=I_2+Y_2$. At the equilibrium $E^{2}_2$, it is valid that
\begin{eqnarray}\label{eq:subsystem2-equilibrium}
    &&(1-v)\Lambda-\frac{\beta_2J_2^*S^*}{N}-\mu S^*=0\nonumber\\
    &&v\Lambda-(\theta_{v2}+\mu)V^*=0\nonumber\\
    &&\frac{\beta_2J_2^*S^*}{N}+\frac{\alpha_{v2}\beta_2J_2^*R_{v1}^*}{N}-(\gamma_2+\mu)J_2^*=0\nonumber\\
    &&\theta_{v2}V^*-\frac{\alpha_{v2}\beta_2J_2^*R_{v1}^*}{N}-\mu R_{v1}^*=0.
\end{eqnarray}

Define the Lyapunov function
\begin{equation}
    L(t)=\left(S-S^*-S^*\ln{\frac{S}{S^*}}\right)+\left(V-V^*-V^*\ln{\frac{V}{V^*}}\right)+\left(J_2-J_2^*-J_2^*\ln{\frac{J_2}{J_2^*}}\right)+\left(R_{v1}-R_{v1}^*-R_{v1}^*\ln{\frac{R_{v1}}{R_{v1}^*}}\right)\nonumber
\end{equation}
in $G=\{(S,V,I_2,C_2,R_2,R_{v1},Y_2,R_{12})\in \Gamma_2;S>0,V>0,J_2>0,R_{v1}>0\}$.

Differentiating $L$ along of the solution of \eqref{eq:equations-only2} and using the equations \eqref{eq:subsystem2-equilibrium}, we have
\begin{eqnarray}
    L'(t)&=&(S-S^*)\left[(1-v)\Lambda\left(\frac{1}{S}-\frac{1}{S^*}\right)-\frac{\beta_2(J_2-J_2^*)}{N}\right]+v\Lambda(V-V^*)\left(\frac{1}{V}-\frac{1}{V^*}\right)\nonumber\\
    &&+(J_2-J_2^*)\left[\frac{\beta_2(S-S^*)}{N}+\frac{\alpha_{v2}\beta_2(R_{v1}-R_{v1}^*)}{N}\right]\nonumber \\
    &&+(R_{v1}-R_{v1}^*)\left[\theta_{v2}\left(\frac{V}{R_{v1}}-\frac{V^*}{R_{v1}^*}\right)-\frac{\alpha_{v2}\beta_2(J_2-J_2^*)}{N}\right]\nonumber\\
    &=&(1-v)\Lambda(S-S^*)\left(\frac{1}{S}-\frac{1}{S^*}\right)+v\Lambda(V-V^*)\left(\frac{1}{V}-\frac{1}{V^*}\right)+\theta_{v2}(R_{v1}-R_{v1}^*)\left(\frac{V}{R_{v1}}-\frac{V^*}{R_{v1}^*}\right). \label{eq:expression-dLdt}
\end{eqnarray}

After some calculations, as in Theorem \ref{theo:subsystem-dfe-2}, we concluded that the expression in \eqref{eq:expression-dLdt}, obtained for $L'(t)$, is non-positive. Furthermore, $L'(t)=0$ if and only if $S=S^*$, $V=V^*$ and $R_{v1}=R_{v1}^*$. Thus,
\begin{eqnarray}
    E&=&\{(S,V,I_2,C_2,R_2,R_{v1},Y_2,R_{12})\in G;L'(t)=0\}\nonumber\\
    &=&\{(S,V,I_2,C_2,R_2,R_{v1},Y_2,R_{12})\in G;S=S^*,V=V^*,R_{v1}=R_{v1}^*\}.\nonumber
\end{eqnarray}
The maximum invariant set of \eqref{eq:equations-only2} contained on the set $E$ is the singleton $\{E^{2}_2\}$, then the endemic equilibrium $E^{2}_2$ is globally asymptotically stable in $G$, by LaSalle’s Invariable Principle \cite{chemostat}. From the Proposition \ref{prop:well-defined}, all orbit of the system \eqref{eq:equations-only2} starting at a point in $\Gamma_2$, with $J_2=I_2+Y_2>0$, belongs to $G$ for $t>0$. Thus, the equilibrium $E^{2}_2$ is globally asymptotically stable in $\{(S,V,I_2,C_2,R_2,R_{v1},Y_2,R_{12})\in \Gamma_2;I_2+Y_2>0\}$.
\end{proof}

\subsection{Global stability}
Next, we will establish conditions for global stability of the DFE.

\begin{lemma}\label{lemma:dfe-global}
Suppose $J_1(0)>0$. Denote $x=(S,V,I_1,I_2,C_1,C_2,R_1,R_2,R_{v1},Y_1,Y_2,R_{12})\in \Gamma$. Denote $\Sigma(t)=S(t)+I_1(t)+I_2(t)+C_2(t)+Y_1(t)+R_2(t)$ for $t\geq 0$. Every orbit of \eqref{eq:equations} in $\Gamma$ enters in
\begin{equation}\label{eq:set-H}
    H=\left\{x\in \Gamma;\Sigma\leq \dfrac{(1-v)\Lambda}{\mu}\right\},
\end{equation}
and $H$ is positively invariant under the flow of \eqref{eq:equations}.
\end{lemma}
\begin{proof}
From the equations of the system, we have
\begin{eqnarray}
    \Sigma'(t)&= & (1-v)\Lambda - \mu \Sigma(t)-\gamma_1 J_1(t).\nonumber
\end{eqnarray}

Using the Comparison Theorem (Theorem B.$1$, \cite{chemostat}), $\Sigma(t)\leq \dfrac{(1-v)\Lambda}{\mu}$ for all $t>0$, if
\begin{equation}
    \Sigma(0)=S(0)+I_1(0)+I_2(0)+C_2(0)+Y_1(0)+R_2(0)\leq \frac{(1-v)\Lambda}{\mu}.\nonumber
\end{equation}
Which implies that $H$ is positively invariant under the flow of \eqref{eq:equations}.

If $J_1(0)>0$, it follows from the equations \eqref{eq:equations} that $J_1(t)>0$ for all $t>0$. If $J_1>0$ and $\Sigma\geq \dfrac{(1-v)\Lambda}{\mu}$, then $\Sigma'<0$. Thus, every forward orbit enters into $H$ after a certain time. 
\end{proof}

With this lemma, we will show the asymptotic stability of the DFE in $H$.

\begin{theorem}\label{theo:DFE-global}
Suppose $\mathcal{R}_0\leq 1$ and $\alpha_1\leq\dfrac{1}{\mathcal{R}_1}$. Let $H$ be as defined in \eqref{eq:set-H}. The DFE, $E^0$, is globally asymptotically stable in $H$.
\end{theorem}  

\begin{proof}
Let $L$ be the Lyapunov function, defined in $H$, by $L=J_1$. Differentiating $L$, with respect to $t$, along of solutions of the model, we have
\begin{eqnarray}
    L'(t)&=&J_1'(t)\nonumber\\
    &=&J_1(\gamma_1+\mu)\left[\frac{\beta_1S}{(\gamma_1+\mu)N}+\frac{\alpha_1\beta_1R_2}{(\gamma_1+\mu)N}-1\right].\nonumber
\end{eqnarray}
Suppose that $\alpha_1\leq 1$. In this case,
\begin{eqnarray}\label{eq:expression3}
    \frac{\beta_1S}{(\gamma_1+\mu)N}+\frac{\alpha_1\beta_1R_2}{(\gamma_1+\mu)N}-1\leq \frac{\beta_1S}{(\gamma_1+\mu)N}+\frac{\beta_1R_2}{(\gamma_1+\mu)N}-1
    =\mathcal{R}_1\frac{S+R_2}{S^0}-1.
\end{eqnarray}
Suppose that $\alpha_1> 1$. In this case,
\begin{eqnarray}\label{eq:expression4}
     \frac{\beta_1S}{(\gamma_1+\mu)N}+\frac{\alpha_1\beta_1R_2}{(\gamma_1+\mu)N}-1\leq \frac{\alpha_1\beta_1S}{(\gamma_1+\mu)N}+\frac{\alpha_1\beta_1R_2}{(\gamma_1+\mu)N}-1= \alpha_1\mathcal{R}_1\frac{S+R_2}{S^0}-1.
\end{eqnarray}
In the set $H$ is valid $S+I_1+I_2+C_2+Y_1+R_2\leq \frac{(1-v)\Lambda}{\mu}=S^0$. Note that, if $J_1=I_1+Y_1>0$, then $S+R_2<S^0$. Thus, using the hypothesis, in both cases, if $J_1>0$, the expressions \eqref{eq:expression3} and \eqref{eq:expression4} are negative. It follows that $L'(t)\leq 0$, and $L'(t)=0$ if and only if $J_1=0$.

Denote $M$ the largest invariant set contained in 
\begin{eqnarray}
    E&=&\{(S,V,I_1,I_2,C_1,C_2,R_1,R_2,R_{v1},Y_1,Y_2,R_{12})\in H;L'(t)=0\}\nonumber\\
    &=&\{(S,V,I_1,I_2,C_1,C_2,R_1,R_2,R_{v1},Y_1,Y_2,R_{12})\in H;J_1=0\}.\nonumber
\end{eqnarray}

It is easy to see that if $J_1=0$, then $C_1$ and $R_1$ tend to zero, when $t$ tends to infinity. Thus, 
$$M \subseteq \{(S,V,I_1,I_2,C_1,C_2,R_1,R_2,R_{v1},Y_1,Y_2,R_{12})\in \Gamma; I_1=C_1=R_1=Y_1=0\}.$$ It follows, from the Theorem \ref{theo:subsystem-dfe-2}, that $M=\{E^0\}$. Thus, from the LaSalle's Invariance Principle \cite{lasalle}, the DFE is asymptotically stable.
\end{proof}

Next, we will give other conditions for the global stability of the DFE.

From the equation, $V'=v\Lambda-(\theta_{v2}+\mu)V$, we have \[\lim_{t\rightarrow +\infty}V(t)=\frac{v\Lambda}{\theta_{v2}+\mu}=V^0.\] It is clear that if $V(0)=V^0$, then $V(t)=V^0$ for all $t\geq0$.

\begin{lemma}\label{lemma:dfe-global2}
Suppose $J_2(0)>0$ and $V(0)=V^0$. Denote $x=(S,V,I_1,I_2,C_1,C_2,R_1,R_2,R_{v1},Y_1,Y_2,R_{12})\in \Gamma$. Denote $\Sigma(t)=S(t)+I_1(t)+I_2(t)+C_1(t)+R_1(t)$ for $t\geq 0$. Every orbit of \eqref{eq:equations} in $\Gamma$ enters in
\begin{equation}\label{eq:set-H2}
    H=\left\{ x\in \Gamma;R_{v1}\leq \dfrac{\theta_{v2}v\Lambda}{\mu(\theta_{v2}+\mu)} \textrm{ and }\Sigma\leq \dfrac{(1-v)\Lambda}{\mu}\right\},
\end{equation}
and $H$ is positively invariant under the flow of \eqref{eq:equations}.
\end{lemma}

\begin{proof}
Supposing the initial condition $V(0)=V^0$ for the variable $V$,
$$R_{v1}'=\theta_{v2}V^0-\alpha_{v2}\beta_2J_2\frac{R_{v1}}{N}-\mu R_{v1}.$$

If $J_2(0)>0$, it follows from the equations \eqref{eq:equations} for $dI_2/dt$ and $dY_2/dt$ that $J_2>0$ for $t>0$. If $J_2>0$ and $R_{v1}\geq \theta_{v2}v\Lambda/\mu(\theta_{v2}+\mu)$,
then $R_{v1}'<0$ and $R_{v1}$ decreases until a value smaller than $\theta_{v2}v\Lambda/\mu(\theta_{v2}+\mu)$.

From the equations of the system, we have
\begin{eqnarray}
    \Sigma'(t)&= & (1-v)\Lambda - \mu \Sigma(t)-\gamma_2I_2-\frac{\beta_2\alpha_2J_2R_1}{N}.\nonumber
\end{eqnarray}

If $J_2(0)>0$, then $I_2>0$ for $t>0$. Thus, if $J_2>0$ and $\Sigma\geq \dfrac{(1-v)\Lambda}{\mu}$, then $\Sigma'<0$ and $\Sigma$ decreases until a value smaller than $\dfrac{(1-v)\Lambda}{\mu}$.

Therefore, every forward orbit of \eqref{eq:equations} enters into $H$ after a certain time.

Using the Comparison Theorem, $R_{v1}(t)\leq \theta_{v2}v\Lambda/\mu(\theta_{v2}+\mu)$ for all $t>0$, if
$R_{v1}(0)\leq \dfrac{\theta_{v2}v\Lambda}{\mu(\theta_{v2}+\mu)}.$
In the same way, $\Sigma(t)\leq \dfrac{(1-v)\Lambda}{\mu}$ for all $t>0$, if
\begin{equation}
    \Sigma(0)=S(0)+I_1(0)+I_2(0)+C_1(0)+R_1(0)\leq \frac{(1-v)\Lambda}{\mu}.\nonumber
\end{equation}
Thus, $H$ is positively invariant under the flow of \eqref{eq:equations}. 
\end{proof}
 
Next, we will show the global stability of the DFE in the set $H$, defined in the previous Lemma.

\begin{theorem}\label{theo:DFE-global2}
Suppose $\mathcal{R}_0\leq 1$ and $\alpha_2\leq\dfrac{1}{\mathcal{R}_2}$. Suppose also $V(0)=V^0$ and $H$ as defined in \eqref{eq:set-H2}. The orbits of \eqref{eq:equations} in $H$ converge for the DFE, $E^0$.
\end{theorem}  

\begin{proof}
Let $L$ be the Lyapunov function, defined in $H$, by $L=J_2$. Differentiating $L$, with respect to $t$, along of solutions of the model, we have
\begin{eqnarray}
    L'(t)&=&J_2'(t)\nonumber\\
    &=&J_2(\gamma_2+\mu)\left[\frac{\beta_2S}{(\gamma_2+\mu)N}+\frac{\alpha_2\beta_2R_1}{(\gamma_2+\mu)N}+\frac{\alpha_{v2}\beta_2R_{v1}}{(\gamma_2+\mu)N}-1\right].\nonumber
\end{eqnarray}

If $\alpha_2\leq 1$, then
\begin{eqnarray}\label{eq:expression5}
\frac{\beta_2S}{(\gamma_2+\mu)N}+\frac{\alpha_2\beta_2R_1}{(\gamma_2+\mu)N}+\frac{\alpha_{v2}\beta_2R_{v1}}{(\gamma_2+\mu)N}-1&\leq&\frac{\beta_2S}{(\gamma_2+\mu)N}+\frac{\beta_2R_1}{(\gamma_2+\mu)N}+\frac{\alpha_{v2}\beta_2R_{v1}}{(\gamma_2+\mu)N}-1\nonumber\\
    &=&\frac{\beta_2(S+R_1-S^0)}{(\gamma_2+\mu)N}+\frac{\alpha_{v2}\beta_2(R_{v1}-R_{v1}^0)}{(\gamma_2+\mu)N}+\mathcal{R}_2-1.
\end{eqnarray}

If $\alpha_2>1$, then
\begin{eqnarray}\label{eq:expression6}
    \frac{\beta_2S}{(\gamma_2+\mu)N}+\frac{\alpha_2\beta_2R_1}{(\gamma_2+\mu)N}+\frac{\alpha_{v2}\beta_2R_{v1}}{(\gamma_2+\mu)N}-1&\leq&\frac{\alpha_2\beta_2S}{(\gamma_2+\mu)N}+\frac{\alpha_2\beta_2R_1}{(\gamma_2+\mu)N}+\frac{\alpha_{v2}\beta_2R_{v1}}{(\gamma_2+\mu)N}-1\nonumber\\
    &\leq&\frac{\alpha_2\beta_2(S+R_1-S^0)}{(\gamma_2+\mu)N}+\frac{\alpha_{v2}\beta_2(R_{v1}-R_{v1}^0)}{(\gamma_2+\mu)N}+\alpha_2\mathcal{R}_2-1.
\end{eqnarray}

Using the hypothesis, we have $\mathcal{R}_2-1\leq0$ and $\alpha_2\mathcal{R}_2-1\leq0$.

In the set $H$ it is valid $R_{v1}\leq R_{v1}^0$, then 
$$\frac{\alpha_{v2}\beta_2(R_{v_1}-R_{v1}^0)}{(\gamma_2+\mu)N}\leq0.$$

Furthermore, it is valid that $\Sigma=S+I_1+I_2+C_1+R_1\leq S^0$. If $J_2>0$, if follows from the equation of the system for $dI_2/dt$, that $I_2(t)>0$, and, therefore, $S+R_1<\Sigma$. Thus, if $J_2>0$, then $S+R_1< S^0$. 

In both cases, we concluded that if $J_2>0$, the expressions \eqref{eq:expression5} and \eqref{eq:expression6} are negative. It follows that $L'(t)\leq 0$, and $L'(t)=0$ if and only if $J_2=0$.

Denote $M$ the largest invariant set contained in 
\begin{eqnarray}
    E&=&\{(S,V,I_1,I_2,C_1,C_2,R_1,R_2,R_{v1},Y_1,Y_2,R_{12})\in H;L'(t)=0\}\nonumber\\
    &=&\{(S,V,I_1,I_2,C_1,C_2,R_1,R_2,R_{v1},Y_1,Y_2,R_{12})\in H;J_2=0\}.\nonumber
\end{eqnarray}

It is easy to see that if $J_2=0$, then $C_2$, $R_2$, $Y_1$ and $R_{12}$ tend to zero, when $t$ tends to infinity. Thus, 
$$M \subseteq \{(S,V,I_1,I_2,C_1,C_2,R_1,R_2,R_{v1},Y_1,Y_2,R_{12})\in \Gamma;I_2=C_2=R_2=Y_1=Y_2=R_{12}=0\}.$$ It follows, from Theorem \ref{theo:subsystem-dfe-1}, that $M=\{E^0\}$. From the LaSalle's Invariance Principle \cite{lasalle}, the DFE is asymptotically stable.
\end{proof}

In resume, we obtained the following theorem about the global stability of the DFE:
\begin{theorem}
Suppose $\mathcal{R}_0\leq 1$. Suppose also $\alpha_1\leq\dfrac{1}{\mathcal{R}_1}$ or $\alpha_2\leq\dfrac{1}{\mathcal{R}_2}$. The DFE, $E^0$, is globally asymptotically stable in $\Gamma$.
\end{theorem}
\begin{proof}
If $\mathcal{R}_0\leq1$ and $\alpha_1\leq 1/\mathcal{R}_1$, the result follows from the Lemma \ref{lemma:dfe-global} and Theorem \ref{theo:DFE-global}. If $\mathcal{R}_0\leq1$ and $\alpha_2\leq 1/\mathcal{R}_2$, it follows from the Lemma \ref{lemma:dfe-global2} and Theorem \ref{theo:DFE-global2}. Note that, in Theorem \ref{theo:DFE-global2}, the omega limit set was assumed to lie in a restricted set (where $V=V^0$), and the equations were analyzed on that set. Here, since that $E^0$ is globally asymptotically stable in this set, we conclude that the asymptotic behavior of the original system is the same (see,  for example, Appendix F \cite{chemostat} about this topic).
\end{proof}

Next, we obtain conditions for the global stability of the boundary endemic equilibria.

\begin{theorem}
Let $J_1(0)>0$. Suppose $\mathcal{R}_2\leq 1$, $\alpha_2\mathcal{R}_2\leq 1$ and $\mathcal{R}_1>1$. Then, the solution tends to endemic equilibrium $E^1$.
\end{theorem}
\begin{proof}
    Suppose $\mathcal{R}_2\leq1$ and $\alpha_2\mathcal{R}_2\leq 1$. Taking the Lyapunov function $L=J_2$ and following the same ideas in Theorem \ref{theo:DFE-global2}, the solution of system tends to the invariant set $M$, where $I_2=C_2=R_2=Y_1=Y_2=R_{12}=0$. Since $\mathcal{R}_1>1$, it follows from Theorem \ref{theo:stability-E1-subsystem} that $M=\{E^1\}$. Thus, from the LaSalle's Invariance Principle, the solution tends to $E^1$.
\end{proof}

\begin{theorem}
Let $J_2(0)>0$. Suppose $\mathcal{R}_1\leq 1$, $\alpha_1\mathcal{R}_1\leq 1$ and $\mathcal{R}_2>1$. Then, the solution tends to endemic equilibrium $E^2$. 
\end{theorem}
\begin{proof}
    This proof is analogous to previous theorem. Just follow initially the ideas of Theorem \ref{theo:DFE-global} and, then, use Theorem \ref{theo:stability-E2-subsystem}.
\end{proof}

\begin{remark}
    Remember that we saw in Theorem \ref{teo:rinv-21} that if $\mathcal{R}_2<1$, $\mathcal{R}_1>1$ and $\mathcal{R}_1^2<1$, then the equilibrium $E^1$ is locally asymptotically stable. It is important to note that the conditions $\mathcal{R}_2<1$ and $\alpha_2\mathcal{R}_2<1$ imply $\mathcal{R}_1^2<1$. In the same way, by Theorem \ref{teo:rinv-12}, if $\mathcal{R}_1<1$, $\mathcal{R}_2>1$ and $\mathcal{R}_2^1<1$, then the equilibrium $E^2$ is locally asymptotically stable. The conditions $\mathcal{R}_1<1$ and $\alpha_1\mathcal{R}_1<1$ imply $\mathcal{R}_2^1<1$. The calculations are showed in Appendix \ref{ap:alfa-rinv}.
\end{remark}

\subsection{Uniform persistence}
Here, based on previous results, we will find conditions to ensure the uniform persistence of the system. We will use the classical Theorem of persistence (Theorem $4.3$ in \cite{persistence}), also used in \cite{li-shuai,dengue-zika-vacina}.

In the following, denote the boundary and the interior of $\Gamma$ as $\partial\Gamma$ and $\Breve{\Gamma}$, respectively. 

\begin{definition}
The system $x'=f(t,x)$ is uniformly persistent, if there is a positive constant $\epsilon$, such as
\begin{equation}
    \liminf _{t \rightarrow \infty}x_i(t)\geq \epsilon, \textrm{  }i=1,...,n,\nonumber
\end{equation}
for all trajectory with positive initial conditions, that is, $x_i(0)>0, i=1,...,n$.
\end{definition}

\begin{theorem}\label{theo:persistence}
Suppose $\mathcal{R}_1>1$, $\mathcal{R}_2>1$, $\mathcal{R}_1^2>1$ and $\mathcal{R}_2^1>1$. The system \eqref{eq:equations} is uniformly persistent in $\Breve{\Gamma}$.  
\end{theorem}

To prove the Theorem, we will use the next Lemmas. 

\begin{lemma}\label{lemma:set-invariants}
Suppose $\mathcal{R}_1>1$ and $\mathcal{R}_2>1$. The largest positively invariant set under the flow of \eqref{eq:equations}, contained
in $\partial\Gamma$, is $\{E^0\}\cup\{E^1\}\cup\{E^2\}$.
\end{lemma}
\begin{proof}
Let $M_{\partial}$, $M_{\partial0}$, $M_{\partial1}$ and $M_{\partial2}$ be the sets
\begin{eqnarray}
    M_{\partial}&=&\{x(0);x(t)\in \partial\Gamma \quad \forall t\geq 0\},\nonumber\\
    M_{\partial0}&=&\{(S,V,I_1,I_2,C_1,C_2,R_1,R_2,R_{v1},Y_1,Y_2,R_{12}) \in \Gamma; J_1=J_2=0\},\nonumber\\
    M_{\partial1}&=&\{ (S,V,I_1,I_2,C_1,C_2,R_1,R_2,R_{v1},Y_1,Y_2,R_{12})\in \Gamma; J_1>0 \textrm{ and } J_2=0\},\nonumber\\
    M_{\partial2}&=&\{ (S,V,I_1,I_2,C_1,C_2,R_1,R_2,R_{v1},Y_1,Y_2,R_{12})\in \Gamma; J_1=0 \textrm{ and } J_2>0\}.\nonumber
\end{eqnarray}

It follows from the system \eqref{eq:equations} that
\begin{eqnarray}
    J_1(t)&=&J_1(0)e^{-(\gamma_1+\mu)t}e^{\int_0^t[\beta_1(S(s)+\alpha_1R_2(s))/N]ds}\nonumber\\
    J_2(t)&=&J_2(0)e^{-(\gamma_2+\mu)t}e^{\int_0^t[\beta_1(S+\alpha_2R_1+\alpha_{v2}R_{v1})/N]ds}.\nonumber    
\end{eqnarray}

For $i=1,2$, it is clear that if $J_i(0)=0$, then $J_i(t)=0$ for all $t>0$. In the same way, if $J_i(0)>0$, then $J_i(t)>0$ for all $t>0$. Thus, $M_{\partial0}$, $M_{\partial1}$ and $M_{\partial2}$ are invariant under the flow of \eqref{eq:equations}.

Here, it is clear that $M_{\partial0}\cup M_{\partial1}\cup M_{\partial2}\subseteq M_\partial$. We will show now that $M_{\partial}\subseteq M_{\partial0}\cup M_{\partial1}\cup M_{\partial2}$.

Suppose $x(0)\in M_\partial$. If $x(0)$ has coordinates satisfying $J_1(0)=J_2(0)=0$, then $x(0) \in M_{\partial0}$. If $x(0)$ has coordinates satisfying $J_1(0)>0$ and $J_2(0)=0$, then $x(0) \in M_{\partial1}$. If $x(0)$ has coordinates satisfying $J_1(0)=0$ and $J_2(0)>0$, then $x(0) \in M_{\partial2}$.

Finally, suppose $x(0)\in M_\partial$ with coordinates satisfying $J_1(0)>0$ and $J_2(0)>0$. From Proposition \ref{prop:well-defined}, $S(t)>0$, $V(t)>0$ and $R_{v1}(t)>0$ for $t>0$. 
Note that, for $i=1,2$,
\begin{equation}\nonumber
    I_i(t)=I_i(0)e^{-(\gamma_i+\mu)t}+\int_0^t\frac{\beta_iJ_i(s)S(s)}{N}e^{-(\gamma_i+\mu)}ds.
\end{equation}
Since that $J_i(0)>0$, then $J_i(t)>0$ for $t\geq 0$. As $S(t)>0$ for all $t$, then $I_i(t)>0$ for $t>0$.      
Now, we will observe the equations for $C_i(t)$ and $R_i(t)$:
\begin{eqnarray}
    C_i(t)&=&C_i(0)e^{-(\theta_j+\mu)t}+\int_0^t\gamma_iI_i(s)e^{-(\theta_j+\mu)(t-s)}ds\nonumber\\
    R_i(t)&=&R_i(0)e^{-\int_0^t\left[\frac{\alpha_j\beta_jJ_j}{N}+\mu\right]ds}+\int_0^t\theta_jC_i(s)e^{-\int_s^t\left[\frac{\alpha_j\beta_jJ_j}{N}+\mu\right]du}ds,\nonumber
\end{eqnarray}
for $i,j\in\{1,2\}$, $i\neq j$.
As $I_i(t)>0$ for $t>0$, then $C_i(t)>0$ for $t>0$. As $C_i(t)>0$ for $t>0$, then $R_i(t)>0$ for $t>0$.

Lastly, we have
\begin{eqnarray}
      Y_1(t)&=&Y_1(0)e^{-(\gamma_1+\mu)t}+\int_0^t\frac{\alpha_1\beta_1J_1(s)R_2(s)}{N}e^{-(\gamma_1+\mu)(t-s)}ds\nonumber\\
   Y_2(t)&=&Y_2(0)e^{-(\gamma_2+\mu)t}+\int_0^t\left(\frac{\alpha_2\beta_2J_2(s)R_1(s)}{N}+\frac{\alpha_{v2}\beta_2J_2(s)R_{v1}(s)}{N}\right)e^{-(\gamma_2+\mu)(t-s)}ds\nonumber\\
   R_{12}(t)&=&R_{12}(0)e^{-\mu t}+\int_0^t\left(\gamma_1Y_1(s)+\gamma_2Y_2(s)\right)e^{-\mu (t-s)}ds.\nonumber
\end{eqnarray}
Since that $J_1,J_2,R_1,R_2,R_{v1}>0$ for $t>0$, then it follows from above equations that $Y_1,Y_2>0$ for $t>0$, and, therefore, $R_{12}>0$ for $t>0$. 

We concluded that if $x(0)\in M_\partial$ satisfies $J_1(0)>0$ and $J_2(0)>0$, then all coordinates $x_i(t)$ are positive for $t>0$. This implies that $x(t)\notin \partial\Gamma$ for $t>0$. Which is a contradiction, since that $x(0)$ was assumed in $M_\partial$. 

Thus, we proved that $M_{\partial}\subseteq M_{\partial0}\cup M_{\partial1}\cup M_{\partial2}$. What implies $M_{\partial}= M_{\partial0}\cup M_{\partial1}\cup M_{\partial2}$.

Now, note that $E^0$ is globally asymptotically stable in $M_{\partial0}$. 
From Theorems \ref{theo:stability-E1-subsystem} and \ref{theo:stability-E2-subsystem}, since that $\mathcal{R}_1>1$ and $\mathcal{R}_2>1$, then $E^1$ is globally asymptotically stable in $M_{\partial1}$, and $E^2$ is globally asymptotically stable in $M_{\partial2}$. Thus, the largest positively invariant set in $\partial\Gamma$ is $\{E^0\}\cup\{E^1\}\cup\{E^2\}$.
\end{proof}

\begin{lemma}\label{lemma:neighborhoods}
 Suppose $\mathcal{R}_1>1$, $\mathcal{R}_2>1$, $\mathcal{R}^2_1>1$ and $\mathcal{R}^1_2>1$. Then,
 \begin{itemize}
     \item[(i)]there is a neighborhood $V_0$ of $E^0$ such that $J_1'>0$ and $J_2'>0$ for all $x \in V_0\setminus \{E^0\}\cap \Breve{\Gamma}$,
     \item[(ii)]there is a neighborhood $V_1$ of $E^1$ such that $J_2'>0$ for all $x \in V_1\setminus \{E^1\}\cap \Breve{\Gamma}$,
     \item[(iii)]there is a neighborhood $V_2$ of $E^2$ such that $J_1'>0$ for all $x \in V_2\setminus \{E^2\}\cap \Breve{\Gamma}$.
 \end{itemize}
\end{lemma}

\begin{proof}
(i) Suppose $\mathcal{R}_1>1$. Let $\delta_0^1$ be
\begin{equation}
    \delta_0^1=\frac{(\mathcal{R}_1-1)(\gamma_1+\mu)N}{2\beta_1(1+\alpha_1)}>0.\nonumber
\end{equation}

Consider a neighborhood $V_0$ of $E^0$, contained in $\Gamma$, such that for all $x \in V_0$, $||x-E^0||<\delta_0^1$. Thus, we have $|S-S^0|<\delta_0^1$ and $|R_2|=|R_2-R_2^0|<\delta_0^1$.
    
We have
\begin{eqnarray}
    J_1'&=&J_1(\gamma_1+\mu)\left[\frac{\beta_1(S-S^0+\alpha_1R_2)}{(\gamma_1+\mu)N}-1+\mathcal{R}_1\right]\nonumber\\
    &>&J_1(\gamma_1+\mu)\left[\frac{-\delta_0\beta_1(1+\alpha_1)}{(\gamma_1+\mu)N}-1+\mathcal{R}_1\right]\nonumber\\
    &=&J_1(\gamma_1+\mu)\left(\frac{\mathcal{R}_1-1}{2}\right).\nonumber
\end{eqnarray}
Thus, for $x\in V_0\setminus\{E^0\}\cap \Breve{\Gamma}$, we have $J_1'>0$, since that $\mathcal{R}_1>1$ and $J_1>0$ in $\Breve{\Gamma}$.

If $\mathcal{R}_2>1$, then let $\delta_0^2$ be a constant
\begin{equation}
    \delta_0^2=\frac{(\mathcal{R}_2-1)(\gamma_2+\mu)N}{2\beta_2(1+\alpha_2+\alpha_{v2})}>0.\nonumber
\end{equation}
By the same reasoning, for $x\in V_0\setminus\{E^0\}\cap\Breve{\Gamma}$, we have $J_2'>0$.

Thus, just take $\delta_0=\min\{\delta_0^1,\delta_0^2\}$.

(ii) Let $\delta_1$ be
\begin{equation}
    \delta_1=\frac{(\mathcal{R}^2_1-1)(\gamma_2+\mu)N}{2\beta_2(1+\alpha_2+\alpha_{v2})}>0.\nonumber
\end{equation}
Consider a neighborhood $V_1$ of $E^1$, contained in $\Gamma$, such that for all $x \in V_1$, $||x-E^1||<\delta_1$. Thus, we have $|S-S^*|<\delta_1$, $|R_1-R_1^*|<\delta_1$ and $|R_{v1}-R_{v1}^*|<\delta_1$.
    
We have
\begin{eqnarray}
    J_2'&=&J_2(\gamma_2+\mu)\left[\frac{\beta_2(S-S^*+\alpha_2(R_1-R_1^*)+\alpha_{v2}(R_{v1}-R_{v1}^*))}{(\gamma_2+\mu)N}-1+\mathcal{R}_1^2\right]\nonumber\\
    &>&J_2(\gamma_2+\mu)\left[\frac{-\delta_1\beta_2(1+\alpha_2+\alpha_{v2})}{(\gamma_2+\mu)N}-1+\mathcal{R}^2_1\right]\nonumber\\
    &=&J_2(\gamma_2+\mu)\left(\frac{\mathcal{R}^2_1-1}{2}\right).\nonumber
\end{eqnarray}
Thus, for $x\in V_1\setminus\{E^1\}\cap \Breve{\Gamma}$, we have $J_2'>0$, since that $\mathcal{R}_1^2>1$ and $J_2>0$ in $\Breve{\Gamma}$.

(iii) Let $\delta_2$ be
\begin{equation}
    \delta_2=\frac{(\mathcal{R}^1_2-1)(\gamma_1+\mu)N}{2\beta_1(1+\alpha_1)}>0.\nonumber
\end{equation}
Consider a neighborhood $V_2$ of $E^2$, contained in $\Gamma$, such that for all $x \in V_2$, $||x-E^0||<\delta_2$. Thus, we have $|S-S^*|<\delta_2$ and $|R_2-R_2^*|<\delta_2$.
    
We have 
\begin{eqnarray}
    J_1'&=&J_1(\gamma_1+\mu)\left[\frac{\beta_1(S-S^*+\alpha_1(R_2-R_2^*))}{(\gamma_1+\mu)N}-1+\mathcal{R}^1_2\right]\nonumber\\
    &>&J_1(\gamma_1+\mu)\left[\frac{-\delta_2\beta_1(1+\alpha_1)}{(\gamma_1+\mu)N}-1+\mathcal{R}^1_2\right]\nonumber\\
    &=&J_1(\gamma_1+\mu)\left(\frac{\mathcal{R}^1_2-1}{2}\right).\nonumber
\end{eqnarray}
Thus, for $x\in V_2\setminus\{E^2\}\cap \Breve{\Gamma}$, we have $J_1'>0$, since that $\mathcal{R}^1_2>1$ and $J_1>0$ in $\Breve{\Gamma}$.
\end{proof}

\begin{proof}(Theorem \ref{theo:persistence})
We will show that the system \eqref{eq:equations} satisfies Theorem D.$2$ in \cite{chemostat} or Theorem $4.3$ in \cite{persistence}. We already know that $\Gamma$ is positively invariant under the flow of \eqref{eq:equations}. 

From Lemma \ref{lemma:set-invariants}, the largest invariant set in $\partial\Gamma$ is $M=\{E^0\}\cup\{E^1\}\cup\{E^2\}$. Assume that $M$ is its own cover. From the proofs of previous Lemmas, each singleton in $M$ is isolated and $M$ is acyclic. Thus, the Hypothesis H of Theorem is satisfied.

From Lemma \ref{lemma:neighborhoods}, there is a neighborhood of $E^0$, contained in $\Gamma$, such that $J_1'>0$ and $J_2'>0$ for all $x \in V_0\setminus\{E^0\}$. Since that $J_1'+J_2'>0$, one among the coordinates $I_1,Y_1,I_2,Y_2$ will increase, then a solution with initial condition in $V_0\setminus\{E^0\}\cap \Breve{\Gamma}$ goes away from $E^0$. In Lemma \ref{lemma:neighborhoods}, $V_0$ is the ball $B(E^0,\delta_0)$. Let $B_0$ be the open ball $B_0=B(E^0,\delta_0/2)$. Of the same way, a solution with initial condition in $V_1\setminus\{E^1\}\cap \Breve{\Gamma}$ goes away from $E^1$ and a solution with initial condition in $V_2\setminus\{E^2\}\cap \Breve{\Gamma}$ goes away from $E^2$. Also analogously, construct the balls $B_1=B(E^1,\delta_1/2)$ and $B_2=B(E^2,\delta_2/2)$.

Let $\delta$ be the constant $\delta=\min\{\delta_0,\delta_1,\delta_2\}$. A solution with initial condition in $y\in S[\partial\Gamma,\delta]\cap \Breve{\Gamma}$ remains in the interior of the compact $\Gamma\setminus(B_0\cup B_1\cup B_2)$, from some $t(y)>0$. Thus, the flow is point dissipative in $S[\partial\Gamma,\delta]\cap \Breve{\Gamma}$.

It follows that any solution of \eqref{eq:equations}, with initial condition in $\Breve{\Gamma}$, get away from the boundary equilibria.

The prove is concluded by observing that the necessary and sufficient condition for uniform persistence in Theorem D.$2$ is equivalent to instability of $E^0$, $E^1$ and $E^2$. 
\end{proof}

Following the steps of the previous theorem, it is possible to find other conditions for the uniform persistence of the system. Here, we will just state the following theorem:

\begin{theorem}\label{theo:persistence2}
Assume that one of the following assumptions is valid:
\begin{itemize}
    \item [(i)] $\mathcal{R}_1>1$, $\mathcal{R}_2<1$ and $\mathcal{R}_1^2>1$, or
    \item[(ii)] $\mathcal{R}_1<1$, $\mathcal{R}_2>1$ and $\mathcal{R}_2^1>1$.
\end{itemize}
Then, the system \eqref{eq:equations} is uniformly persistent in $\Breve{\Gamma}$.
\end{theorem}

Note that, if hypothesis (i) is valid, then there are only two boundary equilibria, $E^0$ and $E^1$. If hypothesis (ii) is valid, then there are only two boundary equilibria, $E^0$ and $E^2$. As before, the hypothesis imply the instability of these equilibria.

\begin{theorem}
In the hypothesis of the Theorem \ref{theo:persistence} or \ref{theo:persistence2}, the system \eqref{eq:equations}, with initial condition in $\Breve{\Gamma}$, is uniformly persistent, and there is an endemic equilibrium in $\Breve{\Gamma}$.
\end{theorem}
\begin{proof}
We will observe that the hypothesis of Theorem $2.8.6$ in \cite{bathia} are satisfied. We already prove that the system is uniformly persistent in $\Breve{\Gamma}$. From this and from the uniform limitation of the solutions, there is a compact set, $A\subset \Breve{\Gamma}$, which is an attractor for the flow of \eqref{eq:equations}. Namely, for all point $x \in A$, $\epsilon\leq x_i\leq N-\epsilon$. The attraction region of $A$ is $\Breve{\Gamma}$. Thus, from the Theorem $2.8.6$ \cite{bathia}, $A$ contains an equilibrium point. Since that $A\subset \Breve{\Gamma}$, this point is an endemic equilibrium.
\end{proof}

In resume, we concluded that the local dynamics is determined by the thresholds $\mathcal{R}_1$, $\mathcal{R}_2$, $\mathcal{R}_1^2$ and $\mathcal{R}_2^1$. The local stability of the disease-free equilibrium (DFE) and endemic equilibria was determined by the basic and invasion reproductive numbers. For $i=1,2$, we denoted $\mathcal{R}_i^{wv}$, the basic reproductive number of strain $i$, in a model without vaccination; and denoted it $\mathcal{R}_i$, in our model, with vaccination. In addition to basic reproductive number, we used the invasion reproductive numbers $\mathcal{R}_i^j$, $i,j \in \{1,2\}, i\neq j$. We show that if $\mathcal{R}_1<1$ and $\mathcal{R}_2<1$, then DFE is stable. Otherwise, it is unstable. For $i,j \in \{1,2\}$, $i\neq j$, if $\mathcal{R}_i>1$, $\mathcal{R}_j<1$ and $\mathcal{R}_i^j<1$, then the endemic equilibrium $E^i$, which has infections only by the strain $i$, is stable. Otherwise, it is unstable. The proofs of global stability were obtained with stronger conditions. Assuming $\mathcal{R}_i<1$ and $\alpha_i\mathcal{R}_i<1$, for $i\in\{1,2\}$, we proved that the solution tends to a set where the strain $i$ is eradicated, that is, $J_i=0$. We emphasize the fact that this hypothesis implies $\mathcal{R}_i<1$ and $\mathcal{R}_j^i<1$, for $j\in\{1,2\}$, $j\neq i$. If in addition we have $\mathcal{R}_j<1$, we concluded that the DFE is globally asymptotically stable; and if $\mathcal{R}_j>1$, the endemic equilibrium $E^j$ is globally asymptotically stable. Lastly, we show that the system is uniformly persistent if $\mathcal{R}_i>1$, $\mathcal{R}_j<1$ and $\mathcal{R}_i^j>1$, for $i,j \in \{1,2\},i\neq j$; or if $\mathcal{R}_i>1$ and $\mathcal{R}_i^j>1$, for $i,j \in \{1,2\},i\neq j$.

\section{Vaccination rate, temporary cross-immunity and ADE effect}\label{sec:simula}
We know that the objective in the vaccination strategy is reducing the basic reproductive number until a value less than one, so that the number of new infections decreases and the diseases eventually disappear from the population. Depending on the parameters related to the diseases and to vaccine, the vaccination strategy may or may not eradicate one or both diseases. In the following, we consider $\mathcal{R}_1^{wv}>1$.

\subsection{Vaccination strategies}
The basic reproductive numbers of the model with and without vaccination are related as
\begin{eqnarray}
    &&\mathcal{R}_1=\mathcal{R}_1^{wv}(1-v)<1 \Longleftrightarrow v>1-\frac{1}{\mathcal{R}_1^{wv}}\label{eq:conditions1}\\
    &&\mathcal{R}_2=\mathcal{R}_2^{wv}[1+v(K-1)]<1\Longleftrightarrow v(K-1)<\frac{1}{\mathcal{R}_2^{wv}}-1, \label{eq:conditions2}
\end{eqnarray}
where $K=\dfrac{\alpha_{v2}\theta_{v2}}{\theta_{v2}+\mu}.$ Thus, the vaccination is always beneficial on the control of strain $1$. On the control of strain $2$, it may or may not be beneficial, depending on the value of $K=K(\alpha_{v2},\theta_{v2})$.  

\begin{remark}\label{obs:alpha-theta}
    For any vaccination rate $v>0$, $$\dfrac{d\mathcal{R}_2}{d\alpha_{v2}}=\dfrac{v\mathcal{R}_2^{wv}\theta_{v2}}{\theta_{v2}+\mu}>0 \textrm{ and }\dfrac{d\mathcal{R}_2}{d(1/\theta_{v2})}=-\dfrac{v\alpha_{v2}\mathcal{R}_2^{wv}\mu}{(1+\mu/\theta_{v2})^2}<0.$$ Thus, the greater the parameter $\alpha_{v2}$, the greater the basic reproductive number for strain $2$, $\mathcal{R}_2$; the greater the cross-immunity period $1/\theta_{v2}$, the smaller $\mathcal{R}_2$.
    \end{remark}

\begin{remark}\label{obs:piora-melhora}
    If $\alpha_{v2}<1+\mu/\theta_{v2}$ $(K<1)$, the vaccination is beneficial on the control of strain $2$, but if $\alpha_{v2}>1+\mu/\theta_{v2}$ $(K>1)$, the vaccination worsens the control of the strain $2$.
\end{remark}
    
Consider $\mathcal{R}_2^{wv}<1$. If $K\leq 1/\mathcal{R}_2^{wv}\iff \alpha_{v2}\leq (1+\mu/\theta_{v2})(1/\mathcal{R}_2^{wv})$, from the equation \eqref{eq:conditions2}, $\mathcal{R}_2<1$ for any vaccination rate. On the other side, if $K>1/\mathcal{R}_2^{wv}$, the vaccination rate $v$ must satisfy $v<\frac{1/\mathcal{R}_2^{wv}-1}{K-1}$ to ensure the stability of DFE. In this case, combining this inequality with the equation \eqref{eq:conditions1}, $v$ has a lower and upper bound:
$$1-\frac{1}{\mathcal{R}_1^{wv}}<v<\frac{1/\mathcal{R}_2^{wv}-1}{K-1}.$$ Furthermore, we must have $$1-\frac{1}{\mathcal{R}_1^{wv}}<\frac{1/\mathcal{R}_2^{wv}-1}{K-1}\iff \alpha_{v2}<\left(1+\frac{\mu}{\theta_{v2}}\right)\left(1+\frac{1/\mathcal{R}_2^{wv}-1}{1-1/\mathcal{R}_1^{wv}}\right).$$

Consider $\mathcal{R}_2^{wv}\geq 1$. If $K<1/\mathcal{R}_2^{wv}$, in addition to equation \eqref{eq:conditions1}, the vaccination rate $v$ must satisfy $v>\frac{1-1/\mathcal{R}_2^{wv}}{1-K}$. Otherwise, $\mathcal{R}_2>1$ for any value of $v$.

In resume, we have the following Theorem.
\begin{theorem}\label{theo:effect-vaccination}
Consider $\mathcal{R}_1^{wv}>1$ and denote $v_1^*=1-\dfrac{1}{\mathcal{R}_1^{wv}}$, $K=\dfrac{\alpha_{v2}\theta_{v2}}{\theta_{v2}+\mu}$, $v_2^*=\dfrac{1/\mathcal{R}_2^{wv}-1}{K-1}$, $\alpha_1^*=\left(1+\dfrac{\mu}{\theta_{v2}}\right)\dfrac{1}{\mathcal{R}_2^{wv}}$ and $\alpha_2^*=\left(1+\dfrac{\mu}{\theta_{v2}}\right)\left(\dfrac{1/\mathcal{R}_2^{wv}-1}{1-1/\mathcal{R}_1^{wv}}+1\right)$. About the stability of the DFE, we have:\\

\noindent 1. Suppose $\mathcal{R}_2^{wv}<1$.
\begin{itemize}
    \item[(i)] If $\alpha_{v2}\leq \alpha_1^*$, the DFE is stable if the vaccination rate $v$ satisfies $v>v_1^*$.
    \item[(ii)] If $\alpha_1^* < \alpha_{v2}<\alpha_2^*,$ a vaccination rate $v$ satisfying $v_1^*<v<v_2^*$ ensures the stability of the DFE.
    \item[(iii)] If $\alpha_{v2}\geq \alpha_2^*$, then the DFE is unstable regardless of vaccination rate $v$.
\end{itemize}
2. Suppose $\mathcal{R}_2^{wv}\geq 1$.
\begin{itemize}
    \item[(i)] If $\alpha_{v2}<\alpha_1^*$, a vaccination rate $v$ satisfying $v>\max\left\{{v_1^*},v_2^*\right\},$ ensures stability of the DFE.
    \item[(ii)] If $\alpha_{v2}\geq \alpha_1^*$, the DFE is unstable regardless of vaccination rate $v$.
\end{itemize}
\end{theorem}

In the previous Theorem, we show conditions in $\alpha$, $\theta$ and $v$ to ensure $\mathcal{R}_0<1$. In the stability analysis, we saw that the local stability of endemic equilibria depends on the invasion numbers, as well as the system persistence conditions. These numbers, in turn, also depends on parameters $\alpha$, $\theta$ and $v$. The analogous fact was observed in \cite{dengue-zika-vacina}, in a model with only the factor $\alpha$. Next, we performed simulations to illustrate our results as a function of these parameters.

\subsection{Numerical simulations}

The parameter values were chosen to represent the infections by the Zika and dengue viruses and can be seen in Table \ref{tab:parameters-fit}. The transmission rates were calculated to obtain the referenced basic reproductive numbers, from the literature. There are not many estimates for the basic reproductive number of Zika. Based on the references, we chose two values (less and greater than one) to run the simulations. It was assumed $\alpha_1=\alpha_2=\alpha_{v2}=\alpha$ and $\theta_1=\theta_2=\theta_{v2}=\theta$. We also assume $\mathcal{R}_1^{wv}=1.3996>1$. The simulations illustrate the results of Theorems \ref{theo:ro-local}, \ref{teo:rinv-21}, \ref{teo:rinv-12}, \ref{theo:persistence}, \ref{theo:persistence2} and \ref{theo:effect-vaccination}.

\begin{table}[ht]
\centering
\caption{Parameters used in the simulations.}
\begin{tabular}{ccccc}
\textbf{Parameter}&\textbf{Range}&\textbf{Assumed}&\textbf{Dimension}&\textbf{Reference}\\
\hline
$N$&$-$&$2.1\times 10^8$&$Dimensionless$&\cite{indexmundi}\\
$\Lambda$&$-$& $2.1\times10^8\times\frac{1}{75\times 52}$&$week^{-1}$&Calculated\\
$\mu$ &$-$& $\frac{1}{75\times 52}$&$weak^{-1}$&\cite{indexmundi}\\ 
$\beta_1$&$1-8$&$1.4$&$week^{-1}$&\cite{dengue-r0}\\
$\beta_2$&$1-5$&$1.0$ or $1.3$&$week^{-1}$&\cite{dengue-zika-vacina,zika-r0,zika-r0-pacific}\\
$\gamma_1$&$-$&$\frac{7}{7}$&$weak^{-1}$&\cite{dengue-brasil}\\
$\gamma_2$&$-$&$\frac{7}{6}$&$weak^{-1}$&\cite{zika-sinan}\\
$\alpha$&$0-5$&$-$&$Dimensionless$&\cite{ade-feng}\\
$\theta$&$\frac{1}{3\times 52}-\frac{1}{52}$&$-$&$weak^{-1}$&\cite{cross-immunity}\\
\hline
\end{tabular}
\label{tab:parameters-fit}
\end{table}

\subsubsection{Scenario 1}
\begin{figure}[!htb]
\centering
\includegraphics[scale=0.7]{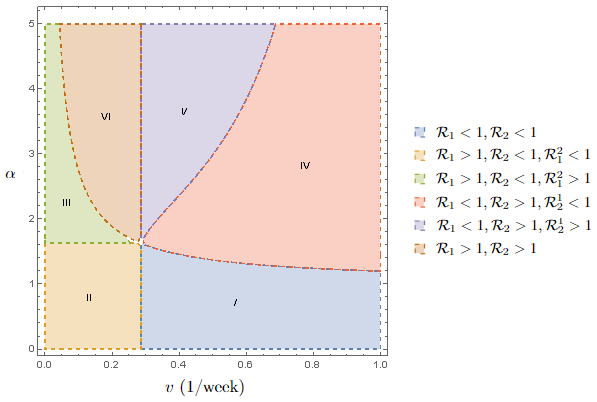}
\quad 
\caption{Basic and invasion reproductive numbers as a function of parameters $\alpha$ and $v$.}
\label{fig:alpha-v}
\end{figure}

In this scenario, we assume $\mathcal{R}_2^{wv}=0.8569<1$. The period of cross-immunity $1/\theta$ is assumed $2$ years. Figure \ref{fig:alpha-v} shows the regions where the invasion and basic reproductive number are greater or less than one, as a function of parameters $\alpha$ and $v$. The curves can see obtained (implicitly or explicitly) from the expressions for $\mathcal{R}_1(v)$, $\mathcal{R}_2(\alpha,v)$, $\mathcal{R}_2^1(\alpha,v)$ and $\mathcal{R}_1^2(\alpha,v)$.

In accordance with Theorem \ref{theo:effect-vaccination} and comments previous to the theorem, for values below $\alpha_1^*$ is possible to obtain $\mathcal{R}_0<1$ if $v>v_1^*$; for values of $\alpha$ between $\alpha_1^*$ and $\alpha_2^*$, if $v$ satisfies $v_1^*<v<v_2^*(\alpha)$, we have $\mathcal{R}_0<1$. Lastly, if $\alpha>\alpha_2^*$, it is not possible to obtain $\mathcal{R}_2<1$ and, therefore, $\mathcal{R}_0>1$ for any value of $v$.

From the theoretical results, in the region $I$, the DFE $E^0$ is stable; in the region $II$, the endemic equilibrium $E^1$ is stable; in the region $IV$, the endemic equilibrium $E^2$ is stable. In the others regions, $III$, $V$ and $VI$, we proved that the strains coexist. Next, we will illustrate this analysis with some examples.

\begin{figure}[!htb]
\centering
\subfloat[]{
\includegraphics[scale=0.7]{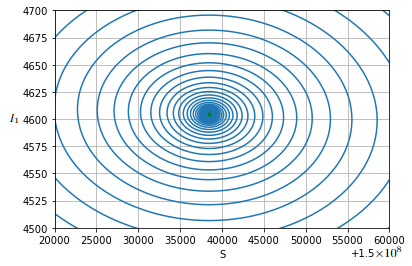}
\label{fig:i-e1-i1}}
\quad 
\subfloat[]{
\includegraphics[scale=0.7]{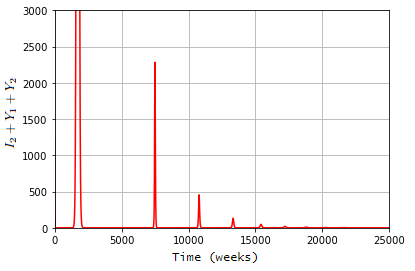}
\label{fig:i-e1-i2}}
\caption{Stability of $E^1$. $\mathcal{R}_1>1$ and $\mathcal{R}_1^2<1$. (a) $S$ and $I_1$ tend to their values at $E^1$. (b) $I_2,Y_1,Y_2$ tend to zero.}
\label{fig:i-e1}
\end{figure}

\begin{figure}[!htb]
\centering
\subfloat[]{
\includegraphics[scale=0.7]{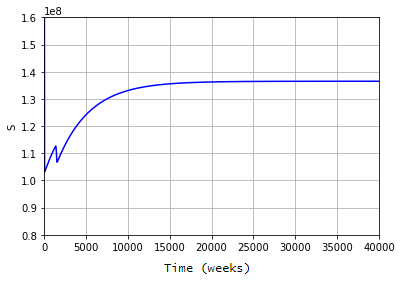}
\label{fig:i-e0-i1}}
\quad 
\subfloat[]{
\includegraphics[scale=0.7]{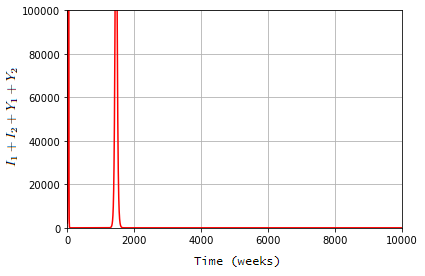}
\label{fig:i-e0-i2}}
\caption{Stability of $E^0$. $\mathcal{R}_1<1$ and $\mathcal{R}_2<1$. (a) $S$ tends to its value at $E^0$ and $I_1$ tends to zero. (b) $I_2,Y_1,Y_2$ tend to zero.}
\label{fig:i-e0}
\end{figure}

\begin{figure}[!htb]
\centering
\subfloat[]{
\includegraphics[scale=0.7]{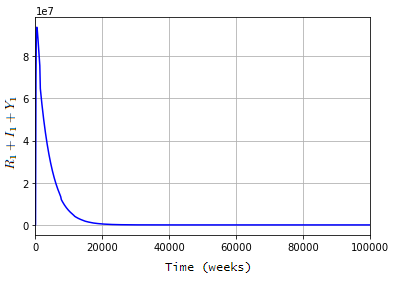}
\label{fig:i-e2-i1}}
\subfloat[]{
\includegraphics[scale=0.7]{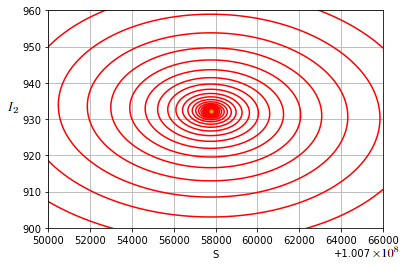}
\label{fig:i-e2-i2}}
 
\subfloat[]{
\includegraphics[scale=0.7]{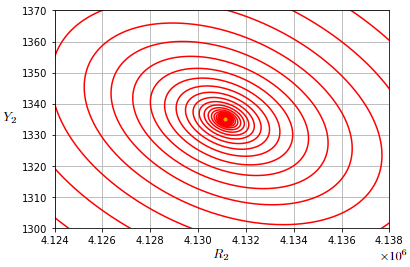}
\label{fig:i-e2-y2}}
\caption{Stability of $E^2$. $\mathcal{R}_2>1$ and $\mathcal{R}_2^1<1$. (a) $I_1,R_1,Y_1$ tend to zero. (b),(c) $S,I_2,R_2,Y_2$ tend to their values at $E^2$.}
\label{fig:i-e2}
\end{figure}

First, suppose $\alpha=1.5$. Note that $\alpha_1^*<\alpha<\alpha_2^*$. We will vary the values of $v$. Figure \ref{fig:i-e1} shows the solution tending to equilibrium $E^1$ for $v=0.2$ $(v<v_1^*)$. In this case, $\mathcal{R}_1=1.1197>1$ and $\mathcal{R}_1^2=0.9697<1$. Figure \ref{fig:i-e0} shows the solution tending to equilibrium $E^0$ for $v=0.35$ $(v_1^*<v<v_2^*)$. In this case, $\mathcal{R}_1=0.9098<1$ and $\mathcal{R}_2=0.9952<1$. The Figure \ref{fig:i-e2} shows the solution tending to equilibrium $E^2$ for $v=0.5$ $(v>v_2^*)$. In this case, $\mathcal{R}_2=1.0545>1$ and $\mathcal{R}_2^1=0.7128<1$. The values of $v=0.2$, $v=0.35$ and $v=0.5$ correspond to regions $II$, $I$ and $IV$, respectively.

\begin{figure}[!htb]
\centering
\subfloat[]{
\includegraphics[scale=0.7]{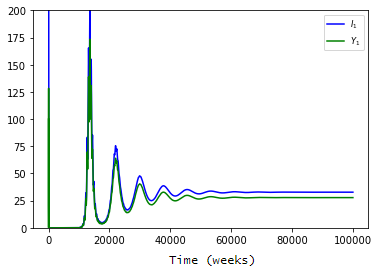}}
\quad 
\subfloat[]{
\includegraphics[scale=0.7]{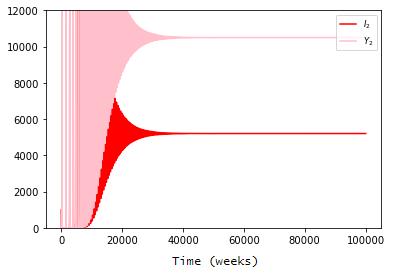}}
\caption{Persistence of both strains. (a) Infections by strain $1$. (b) Infections by strain $2$.}
\label{fig:persistence}
\end{figure}

In accordance with Theorem \ref{theo:effect-vaccination}, these simulations show that the DFE is stable for an intermediary vaccination rate. 
Now, observe the effect of the parameter related to ADE, $\alpha$, for a fixed vaccination rate, $v=0.5$. We know that if $v=0.5$ then $\mathcal{R}_1<1$. From Figure \ref{fig:alpha-v}, for small values of $\alpha$, we have $\mathcal{R}_2<1$ (region $I$). However, as seen, for $\alpha=1.5$ we have $\mathcal{R}_2>1$ and strain $2$ persists (Figure \ref{fig:i-e2}, region $IV$). In this case $\mathcal{R}_2^1<1$. Suppose now $\alpha=3$ (region $V$). Figure \ref{fig:persistence} shows the persistence of both strains. In this case, $\mathcal{R}_1=0.6998<1$, $\mathcal{R}_2=1.6805>1$ and $\mathcal{R}_2^1=1.0031>1$. Here, it is possible to see that although the basic reproductive number of strain $1$ is less than one, its invasion reproductive number is greater than one and it can persist in the population. Note that, according to Theorem \ref{theo:subsystem-dfe-1}, in the absence of strain $2$, strain $1$ would be eradicated. The high value of $\alpha$ cause a synergy between the strains, what allows the coexistence of them.

\subsubsection{Scenario 2}
\begin{figure}[!htb]
\centering
\subfloat[]{
\includegraphics[scale=0.7]{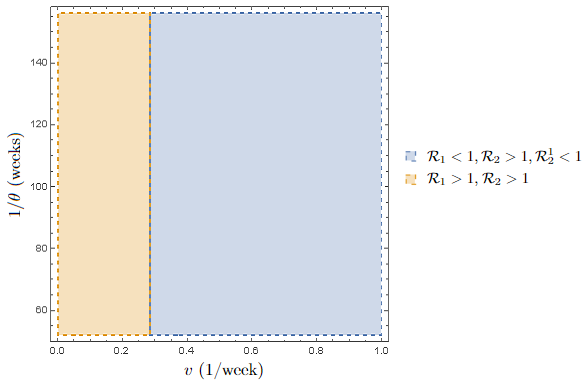}
\label{fig:theta-v}}
\quad
\subfloat[]{
\includegraphics[scale=0.7]{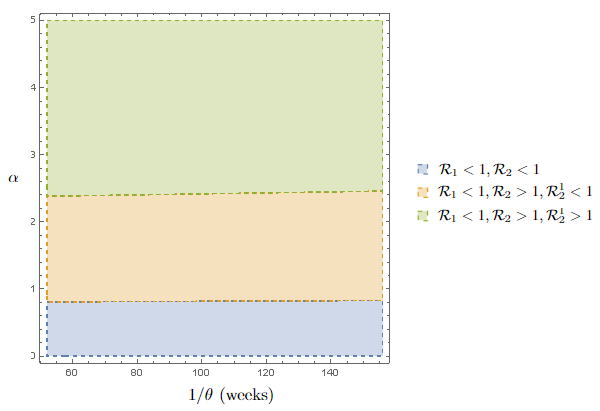}
\label{fig:alpha-theta}}
\caption{Basic and invasion reproductive numbers as a function of the parameters (a) $1/\theta$ and $v$; (b) $\alpha$ and $1/\theta$.}
\end{figure}

Suppose $\mathcal{R}_2^{wv}=1.1140>1$. The parameter related to ADE was assumed $\alpha=1.0$. That is, primary infections do not enhance, nor protect against secondary infections. Figure \ref{fig:theta-v} shows the regions where the invasion and basic reproductive number are greater or less than one, as a function of the parameters $1/\theta$ and $v$. The blue region corresponds to $\mathcal{R}_1<1$, that is $v>v_1^*$. In all this region, we have $\mathcal{R}_2>1$; the invasion and reproductive numbers indicate the persistence of strain $2$. This suggests the persistence of strain $2$ regardless of vaccination rate. Nonetheless, from Remark \ref{obs:piora-melhora}, with the vaccination, we expected a decrease on the number of new infections by strain $2$.

As said in Remark \ref{obs:alpha-theta}, $\frac{d\mathcal{R}_2}{d(1/\theta)}=-\frac{v\alpha_{v2}\mathcal{R}_2^{wv}\mu}{(1+\mu/\theta)^2}<0$. Nonetheless, since that $|\frac{d\mathcal{R}_2}{d(1/\theta)}|<\mathcal{R}_2^{wv}\mu=\frac{\mathcal{R}_2^{wv}}{75\times 52}$, the derivative is negative, but its absolute value is very small, what justifies the variation in the period of cross-immunity has no effect. 

\subsubsection{Scenario 3}
In this last scenario, as before, $\mathcal{R}_2^{wv}=1.1140>1$. Assume $v=0.5$. This vaccination rate is enough to obtain $\mathcal{R}_1<1$. We will analyze if there are values of $\alpha$ and $1/\theta$ such that both diseases can be eradicated. Figure \ref{fig:alpha-theta} shows the invasion and basic reproductive numbers as a function of $\alpha$ and $1/\theta$. It is possible to see that the variation in $\theta$ does not have effect in these reproductive numbers. On the other hand, it is possible to achieve $\mathcal{R}_2<1$ for $\alpha$ below one. Intermediary values of $\alpha$ cause permanence of strain $2$. High values of $\alpha$ cause synergy between the strains. The low vaccination rate allows infections by strain $1$, leading to permanence of both in the population despite $\mathcal{R}_1<1$.  

 From Theorem \ref{theo:effect-vaccination}$-2$, to obtain $\mathcal{R}_2<1$ with some vaccination strategy (for some rate $v$), we must have $\alpha<\alpha_1^*=(1+\mu/\theta)(1/\mathcal{R}_2^{wv})$. For the considered values, $(1+\mu/\theta)\approx 1$. If $(1/\mathcal{R}_2^{wv})<1$, $\alpha_1^*$ is less than one or approximately one.

\section{Discussion}\label{sec:discussão}

Based on the last findings concerning Zika and dengue viruses, we analyzed the possible results of a vaccination strategy against one strain in a two-strain model that takes into account temporary cross-immunity and antibody-dependent enhancement (ADE). When we study vaccination strategies, look for a strategy that reduces the basic reproductive number, $\mathcal{R}_0=\max\{\mathcal{R}_1,\mathcal{R}_2\}$, to a value less than one, expecting that the number of new infections decreases until eventually, the disease disappears from the population. Supposing the vaccination against strain $1$, it is important to note that $\mathcal{R}_2$ is increasing as a function of the factor related to ADE $(\alpha)$, and decreasing as a function of the period of cross-immunity $(1/\theta)$. As we have two strains with some competition and some synergy between them, we expected to reduce $\mathcal{R}_1$, and, if possible, to reduce also $\mathcal{R}_2$. 

In the first moment, we studied the dynamics of the model through the basic and invasion reproductive numbers. It was shown, for example, the local stability of DFE when $\mathcal{R}_0<1$. The asymptotic global stability was proved supposing also $\alpha_i\mathcal{R}_i<1$, for $i=1 \textrm{ or }2$. Note that if there is no ADE ($\alpha\leq 1$), $\mathcal{R}_0<1$ ensures the global stability. We also provide conditions for the stability of the endemic equilibria and the coexistence of strains.

Then, in Theorem \ref{theo:effect-vaccination}, we exhibited the needed vaccination rates to obtain $\mathcal{R}_1<1$ and, when possible, $\mathcal{R}_2<1$, as a function of $\alpha_{v2}$ and $\theta_{v2}$, parameters referring to cross-immunity and ADE, from the vaccination. First, it was assumed $\mathcal{R}_2^{wv}$ (model without vaccination) is less than one. For small values of $\alpha_{v2}$ $(\alpha_{v2}<\alpha_1^*(\theta_{v2}))$, we found a minimum vaccination rate, $v_1^*$, to ensure the stability of DFE. For intermediary values of $\alpha_{v2}$ $(\alpha_1^*(\theta_{v2})<\alpha_{v2}<\alpha_2^*(\theta_{v2}))$, a vaccination rate $v$, $v_1^*<v<v_2^*(\alpha_{v2},\theta_{v2})$, ensures the stability of DFE. Lastly, for high values of $\alpha_{v2}$ $(\alpha_{v2}\geq \alpha_2^*(\alpha_{v2},\theta_{v2}))$, it is not possible to eradicate both diseases. In the worst case, with $\mathcal{R}_2^{wv}>1$, for small values of $\alpha_{v2}$ $(\alpha_{v2}<\alpha_1^*(\theta_{v2}))$, a minimum vaccination rate is required, $v>\max\{v_1^*,v_2^*(\alpha_{v2},\theta_{v2})\}$. For greater values of $\alpha_{v2}$ $(\alpha_{v2}>\alpha_1^*(\alpha_{v2},\theta_{v2}))$, it is not possible to eradicate both diseases. 

Simulations were done supposing that the period of cross-immunity and the level of cross-susceptibility are the same for both strains, and the vaccine has the same effect as an infection. The basic and invasion reproductive numbers were analyzed as a function of $\alpha$, $\theta$ and $v$.

 We simulated a case where $\mathcal{R}_2^{wv}<1$. The results of vaccination strategies can be the persistence of only one of the strains, the coexistence, or the eradication of both. We fixed the period of cross-immunity $(1/\theta)$ equal to 2 years. With the assumed parameters, for $\alpha<\alpha_2^*(\theta)\approx 1.6$, there is a vaccination strategy such that it is possible to ensure the stability of the DFE. Above this value, we have $\mathcal{R}_2>1$, indicating the persistence of strain $2$ or both strains.

We also analyzed a case where $\mathcal{R}_2^{wv}>1$. We assume $\alpha=1$, that is, there is no enhancement nor protection from primary infections, and observe if temporary cross-immunity allows eradication of strain $2$. For a period of cross-immunity as expected ($1$ to $3$ years), it is not possible to obtain $\mathcal{R}_2<1$. This suggests the persistence of strain $2$, regardless of vaccination rate. 

Lastly, with the same $\mathcal{R}_2^{wv}>1$, we fixed the vaccination rate $v=0.5$. This vaccination rate ensures the eradication of strain $1$. We observed if there are values of $\alpha$ and $\theta$ such that the DFE is stable. The results does not vary much with $\theta$. The values of $\alpha$ determine if $\mathcal{R}_0<1$ or $\mathcal{R}_0>1$. Intermediary values of $\alpha$ keep $\mathcal{R}_2>1$. High values of $\alpha$ can cause synergy between the strains with the persistence of strain $1$, despite vaccination ensuring $\mathcal{R}_1<1$. For the considered values, the threshold for $\alpha$, which allows $\mathcal{R}_2<1$, $(\alpha<(1+\mu/\theta)\times1/\mathcal{R}_2^{wv})$, is below one. Note that if the average life expectancy is much greater than the period of cross-immunity $(1/\mu>>1/\theta)$, then $1+\mu/\theta \approx 1$. 

In the case $\mathcal{R}_2^{wv}>1$, even when it is not possible to obtain $\mathcal{R}_2<1$, with the vaccination, we can expect a decrease on the number of new infections by the strain $2$. For this, we must have $\alpha<(1+\mu/\theta)$. With the parameters used, this bound is greater than one, but very close to one. For example, for cross-immunity of $2$ years and average life expectation of $75$ years, this bound is $1.03$. 

These results indicate that the vaccination may or may not be beneficial on the control of strain $2$. If strain $2$ has basic reproductive number less than one, the cross-immunity can contribute to eradication of both strains. If, on the other hand, the reproductive basic number is greater than one, the existence or not of antibody-dependent enhancement can determine the eradication of strain $2$.   

\section{Acknowledgement}
Lorena C. Bulhosa was supported by a grant from the Conselho Nacional de Desenvolvimento Científico e Tecnológico (CNPq) of Brazil [proc. 141180/2017-0]. Juliane F Oliveira was supported by a grant from the Oswaldo Cruz Foundation [grant number VPGDI-050-FIO-20-2-10, 2022]. The funders of the study had no role in the study design, data collection, data analysis, data interpretation, or the writing of the manuscript.

\appendix
\section{Proof of Proposition \ref{prop:well-defined}}
\label{ap:well-defined}
\begin{proof}
Consider $x=(S,V,I_1,I_2,C_1,C_2,R_1,R_2,R_{v1},Y_1,Y_2,R_{12})$ and suppose that $x(0)\geq 0$. Then for any $t>0$, we have
\begin{eqnarray}
   S(t)&=&S(0)e^{-\int_0^t\left(\beta_1J_1(s)/N+\beta_2J_2(s)/N+\mu \right)ds}+(1-v)\Lambda \int_0^te^{-\int_s^t\left(\beta_1J_1(u)/N+\beta_2J_2(u)/N+\mu \right)du}ds \geq 0\nonumber\\
   V(t)&=&\left[V(0)+\frac{\Lambda v}{\theta_{v2}+\mu}\right]e^{-(\theta_{v2}+\mu)t}+\frac{\Lambda v}{\theta_{v2}+\mu} \geq 0\nonumber\\
   R_{v1}(t)&=&R_{v1}(0)e^{-\int_0^t(\alpha_{v2}\beta_2J_2(s)/N+\mu)ds}+\theta_{v2}\int_0^tV(s)e^{-\int_s^t(\alpha_{v2}\beta_2J_2(u)/N+\mu)du}ds \geq 0.\nonumber\\
   J_1(t)&=&J_1(0)e^{-(\gamma_1+\mu)t}e^{\int_0^t[\beta_1(S(s)+\alpha_1R_2(s))]/Nds}\geq 0\nonumber\\
   J_2(t)&=&J_2(0)e^{-(\gamma_2+\mu)t}e^{\int_0^t[\beta_1(S+\alpha_2R_1+\alpha_{v2}R_{v1})/Nds]}\geq 0.\nonumber\\
   I_i(t)&=&I_i(0)e^{-(\gamma_i+\mu)t}+\int_0^t\frac{\beta_i(s)J_i(s)S(s)}{N}e^{-(\gamma_i+\mu)(t-s)}ds\geq 0, \quad i\in\{1,2\}\nonumber\\
   C_i(t)&=&C_i(0)e^{-(\theta_j+\mu)t}+\int_0^t\gamma_iI_i(s)e^{-(\theta_j+\mu)(t-s)}ds\geq 0, \quad i,j\in\{1,2\},i\neq j \nonumber\\
   R_i(t)&=&R_i(0)e^{-\int_0^t(\alpha_j\beta_jJ_j(s)/N+\mu) ds}+\theta_j\int_0^tC_i(s)e^{-\int_s^t(\alpha_j\beta_jJ_j(u)/N+\mu)du}ds\geq 0,\quad  i,j\in\{1,2\},i\neq j\nonumber\\
   Y_1(t)&=&Y_1(0)e^{-(\gamma_1+\mu)t}+\int_0^t\frac{\alpha_1\beta_1J_1(s)R_2(s)}{N}e^{-(\gamma_1+\mu)(t-s)}ds\geq 0\nonumber\\
   Y_2(t)&=&Y_2(0)e^{-(\gamma_2+\mu)t}+\int_0^t\left(\frac{\alpha_2\beta_2J_2(s)R_1(s)}{N}+\frac{\alpha_{v2}\beta_2J_2(s)R_{v1}(s)}{N}\right)e^{-(\gamma_2+\mu)(t-s)}ds\geq 0\nonumber\\
   R_{12}(t)&=&R_{12}(0)e^{-\mu t}+\int_0^t\left(\gamma_1Y_1(s)+\gamma_2Y_2(s)\right)e^{-\mu (t-s)}ds\geq 0.\nonumber
\end{eqnarray}

In particular, if  $S(0)>0$, then $S(t)$, $V(t)$ and $R_{v1}(t)$ are strictly positive for all $t>0$. Thus, the invariance of $\mathbb{R}^{12}_+$ under the flow follows directly from the equations in \eqref{eq:equations}.

Since we supposed that the total population is constant and equal to $\Lambda/\mu$,  together with the invariance of $\mathbb{R}^{12}_+$, we can conclude that solutions are limited. 

Lastly, given an initial condition in $\mathbb{R}^{12}_+$, the existence and uniqueness of solutions follows from the fact that vector field is a continuous and Lipschitz function in $\mathbb{R}^{12}_+$.
\end{proof}

\section{Calculations of Remark \ref{rem:alfa-rinv}}\label{ap:alfa-rinv}

\begin{eqnarray}
\mathcal{R}_1^2&=&\frac{\beta_2S^*}{(\gamma_2+\mu)N}+\frac{\beta_2(\alpha_2R_1^*+\alpha_{v2}R_{v1}^*)}{(\gamma_2+\mu)N}=\frac{\beta_2}{\gamma_2+\mu}\frac{S^*+\alpha_2R_1^*+\alpha_{v2}R_{v1}^*}{N}.\nonumber
\end{eqnarray}

\noindent If $\alpha_2\leq 1$, the above expression is less or equal to
$$\frac{\beta_2}{\gamma_2+\mu}\frac{S^*+R_1^*+\alpha_{v2}R_{v1}^*}{N}.$$
If $\alpha_2>1$, the above expression is less or equal to
$$\frac{\alpha_2\beta_2}{\gamma_2+\mu}\frac{S^*+R_1^*+\alpha_{v2}R_{v1}^*}{N}.$$

We have that
\begin{eqnarray}
    \frac{S^*+R_1^*}{N}&=&\frac{\gamma_1+\mu}{\beta_1}+\frac{\theta_2\gamma_1(1-v)}{(\theta_2+\mu)(\gamma_1+\mu)}\left(1-\frac{1}{\mathcal{R}_1}\right)\nonumber\\
    &=&\frac{\theta_2\gamma_1}{(\theta_2+\mu)(\gamma_1+\mu)}(1-v)-\frac{\gamma_1+\mu}{\beta_1}\left(1-\frac{\theta_2\gamma_1}{(\theta_2+\mu)(\gamma_1+\mu)}\right)\nonumber\\
    &\leq&\frac{\theta_2\gamma_1}{(\theta_2+\mu)(\gamma_1+\mu)}(1-v)\leq 1-v.
\end{eqnarray}
We also have $\dfrac{\alpha_{v2}R_{v1}^*}{N}=\dfrac{\alpha_{v2}\theta_{v2}v}{\theta_{v2}+\mu}.$

Thus,
$$\frac{\beta_2}{\gamma_2+\mu}\frac{S^*+R_1^*+\alpha_{v2}R_{v1}^*}{N}\leq \frac{\beta_2}{\gamma_2+\mu}\left[1-v+\frac{v\alpha_{v2}\theta_{v2}}{\theta_{v2}+\mu}\right]=\mathcal{R}_2.$$
In resume, if $\alpha_2\leq 1$, then $\mathcal{R}_1^2\leq \mathcal{R}_2$; if $\alpha_2>1$, then  $\mathcal{R}_1^2\leq \alpha_2\mathcal{R}_2$.

In the same way, we have
$$\mathcal{R}_2^1=\frac{\beta_1S^*}{(\gamma_1+\mu)N}+\frac{\alpha_1\beta_1R_2^*}{(\gamma_1+\mu)N}=\frac{\beta_1}{\gamma_1+\mu}\frac{S^*+\alpha_1R_2^*}{N}.$$

If $\alpha_1\leq 1$, then the above expression is less or equal to $\frac{\beta_1}{\gamma_1+\mu}\frac{S^*+R_2^*}{N}$. If $\alpha_1>1$, then the above expression is less or equal to $\frac{\alpha_1\beta_1}{\gamma_1+\mu}\frac{S^*+R_2^*}{N}$.

We have that
\begin{eqnarray}
    \frac{S^*+R_2^*}{N}=(1-v)\left[\frac{\mu}{x+\mu}+\frac{x\gamma_2\theta_1}{(x+\mu)(\gamma_2+\mu)(\theta_1+\mu)}\right]\leq(1-v)\left[\frac{\mu}{x+\mu}+\frac{x}{(x+\mu)}\right]=1-v.\nonumber
\end{eqnarray}

Thus, 
$$\frac{\beta_1}{\gamma_1+\mu}\frac{S^*+R_2^*}{N}\leq \frac{\beta_1}{\gamma_1+\mu}(1-v)=\mathcal{R}_1.$$
In resume, if $\alpha_1\leq 1$, then $\mathcal{R}_2^1\leq \mathcal{R}_1$; if $\alpha_1>1$, then  $\mathcal{R}_2^1\leq \alpha_1\mathcal{R}_1$.

\section{Coefficients of Q}\label{ap:coef}
The coefficients $b,c$ and $d$ of $Q(\lambda)$ are given by
\begin{eqnarray}
    b&=&\frac{\beta_2J_2}{N}+\frac{\alpha_{v2}\beta_2J_2}{N}+2\mu\nonumber\\
c&=&\left(\frac{\beta_2J_2}{N}+\mu\right)\left(\frac{\alpha_{v2}\beta_2J_2}{N}+\mu\right)+\frac{\alpha_{v2}^2\beta_2^2R_{v1}J_2}{N^2}+\frac{\beta_2^2SJ_2}{N^2}\nonumber\\
    d&=&\frac{\alpha_{v2}^2\beta_2^2R_{v1}J_2}{N^2}\left(\frac{\beta_2J_2}{N}+\mu\right)+\frac{\beta_2^2SJ_2}{N^2}\left(\frac{\alpha_{v2}\beta_2J_2}{N}+\mu\right)\nonumber.
\end{eqnarray}

\bibliographystyle{plain}
\bibliography{bibliography}

\end{document}